%% file: ms.tex
\documentclass[12pt,preprint]{aastex}
\input{macros}

\shortauthors{Gallimore et al.}
\shorttitle{\h2o\ masers of NGC~1068}
\begin{document}
\singlespace
\title{The Nature of the Nuclear \h2o\ Masers of NGC~1068:
Reverberation and Evidence for a Rotating Disk Geometry}   

\author{J.F. Gallimore\altaffilmark{1,2}, C. Henkel\altaffilmark{3},
S. A. Baum\altaffilmark{4}, I. S. Glass\altaffilmark{5},
M. J. Claussen\altaffilmark{6}, M. A. Prieto\altaffilmark{7}, and
A. von Kap-herr\altaffilmark{3}}

\altaffiltext{1}{Department of Physics, Bucknell University,
Lewisburg, PA 17837, {jgallimo@bucknell.edu}}
\altaffiltext{2}{National Radio Astronomy Observatory, 520 Edgemont Rd, 
Charlottesville, VA 22903}
\altaffiltext{3}{Max-Planck-Institut f\"ur Radioastronomie, Auf dem
H\"ugel 69, D-53121 Bonn, Germany} 
\altaffiltext{4}{Space Telescope Science Institute, 3700 San Martin Dr.,
Baltimore, Maryland 21218}
\altaffiltext{5}{South African Astronomical Observatory, PO Box 9,
Observatory 7935, South Africa}
\altaffiltext{6}{NRAO, P.O. Box O, Socorro, NM 87801-0387}
\altaffiltext{7}{European Southern Observatory,
Karl-Schwarzschild-Strasse 2, D-85740 Garching b. M\"unchen, Germany}


\clearpage

\input{abstract}
\input{intro}

\input{obs}

\input{results}

\input{discussion}

\input{ack}
\clearpage

\input{bib}
\clearpage
\input{fig}
\clearpage
\input{tables}

\end{document}

%% file: macros.tex
\newcommand{\lapprox }{{\lower0.8ex\hbox{$\buildrel <\over\sim$}}}
\newcommand{\gapprox }{{\lower0.8ex\hbox{$\buildrel >\over\sim$}}}
 
\newcommand{\boiiib}{\mbox {[O\,{\sc iii}]}}            
\newcommand{\hi}{\mbox {H\,{\sc i}}}                  

\def\h2o{H$_2$O}
\def\kms{km s$^{-1}$}

\def\mone{$^{-1}$}

\def\mthree{$^{-3}$}
\def\Msun{\hbox{${\rm M}_{\sun}$}}

\def\h2o{H$_2$O}
\def\d21{D_{21}}

\def\mc#1{\multicolumn{1}{c}{#1}}
\def\mcc#1{\multicolumn{2}{c}{#1}}
\def\mccc#1{\multicolumn{3}{c}{#1}}

\def\n11{n_{11}}
\def\n10{n_{10}}
\def\t400{T_{400}}
\def\f1{f_{(-2)}}

\def\e4{\epsilon_{(-4)}}

%% file: abstract.tex
\begin{abstract}
We report new (1995) Very Large Array observations and (1984--1999)
Effelsberg 100m monitoring observations of the 22~GHz \h2o\ maser
spectrum of the Seyfert 2 galaxy NGC~1068. The sensitive VLA
observations provide a registration of the 22~GHz continuum emission
and the location of the maser spots with an accuracy of $\sim 5$~mas.
Within the monitoring data, we find evidence that the nuclear masers
vary coherently on time-scales of months to years, much more rapidly
than the dynamical time-scale. We argue that the nuclear masers are
responding in reverberation to a central power source, presumably the
central engine. Between October and November 1997, we detected a
simultaneous flare of the blue-shifted and red-shifted satellite maser
lines. Reverberation in a rotating disk naturally explains the
simultaneous flaring. There is also evidence that near-infrared
emission from dust grains associated with the maser disk also responds to
the central engine. We present a model in which an X-ray flare results
in both the loss of maser signal in 1990 and the peak of the
near-infrared light-curve in 1994.  In support of rotating disk
geometry for the nuclear masers, we find no evidence for centripetal
accelerations of the redshifted nuclear masers; the limits are $\pm
0.006$~\kms\ year\mone, implying that the masers are located within
2\arcdeg\ of the kinematic line-of-nodes.  We also searched for high
velocity maser emission like that observed in NGC~4258. In both VLA
and Effelsberg spectra, we detect no high velocity lines between $\sim
\pm 350$~\kms-- $\pm 850$~\kms\ relative to systemic, arguing that
masers only lie outside a radius of $\sim 0.6$~pc (1.9 light-years)
from the central engine (assuming a distance of 14.4~Mpc). We also
consider possible models for the jet masers near radio continuum
component~C. We favor a shock-precursor model, in which the molecular
gas surrounding the jet is heated by X-ray emission from a shock front
between the jet and a molecular cloud.
\end{abstract}

%% file: intro.tex
\section{Introduction}\label{intro}

\objectname[]{NGC~1068} is one of the classical Seyfert galaxies
\citep{Seyfert43} and is the prototype for type 2 Seyferts
(\citealt{KW74}). In polarized light, however, the optical spectrum
has characteristics of Seyfert type 1 spectra \citep{AM85}. The model
is that the broad-line region and central engine are obscured by an
edge-on, dusty molecular disk and can be seen only in reflected (and
therefore polarized) light. NGC~1068 is one of the original \h2o\
megamaser galaxies
\citep{CHL84}, and it was proposed early on that these masers may trace
warm molecular gas in the obscuring disk
\citep{CL86}.

NGC~1068 is unusual among megamaser galaxies because there appears to
be two distinct sources of \h2o\ maser emission associated with
compact features of the radio jet. The inner arcsecond of the radio jet of
NGC~1068 resolves into several compact sources connected by fainter,
linear radio structure. For reference, Figure~\ref{continuum} provides
a new 22~GHz image of the sub-arcsecond jet with the principal compact
features NE, C, S1, and S2 labeled. \h2o\ maser emission associates
with components S1 and C \citep{maserpaper}. Discussed further below,
virtually all of the current evidence points to S1 as the location of
the hidden nucleus (e.g., \citealt{naturepaper}). As much for
convenience as interpretation, we therefore refer to the masers
associated with S1 as ``the nuclear masers'' and the masers associated
with component C as the ``jet masers.'' We adopt a distance of
14.4~Mpc \citep{Tully88}, corresponding to a scale of $1\arcsec =
69.8$~pc, or 228~light-years.

\subsection{The Nuclear Masers at Component S1}

The nuclear masers at S1 span a velocity range of roughly 600~\kms\
(see Figures~\ref{bluespec}--\ref{highv}). The brightest nuclear
masers concentrate near three discrete velocities, nearly symmetric
about the systemic velocity of the galaxy ($v_{LSR} = 1137$~\kms;
\citealt{BSTT97}). In the LSR inertial frame and adopting the optical
convention for Doppler velocities, the nuclear masers comprise the
``red masers'' near $cz
\sim 1430$~\kms, the ``systemic masers'' near 1140~\kms, and the
``blue masers'' centered near 860~\kms. The red masers are the
brightest and have enjoyed the most detailed study; in contrast, the
weaker blue masers have been largely ignored, presumably owing to
sensitivity considerations (\citealt{NIMMH95};
\citealt{maserpaper}). As a result, there are very few monitoring
observations of the systemic and blue masers.

Apart from the presence of masers, S1 is peculiar in two
respects. Firstly, the radio continuum spectrum is consistent with
thermal free-free emission turning opaque below $\sim 5$~GHz
(\citealt{n1068b}; \citealt{naturepaper}; \citealt{RCWU98}). Secondly, VLBA observations resolve S1 into a parsec-long,
vaguely linear source aligned at right angles to the local radio jet
axis \citep{naturepaper}. The brightness temperatures over the 5--8~GHz
frequency interval range from $10^5$--$10^6$~K. These results argue that S1
traces thermal emission from a disk rather than synchrotron emission
from a jet. We proposed that S1 might trace free-free emission from
ionized gas on the inner surface of the obscuring torus
(\citealt{n1068b}; \citealt{naturepaper}).

Since the discovery of the blue-shifted masers, VLBA observations have
resolved the position-velocity distribution of the nuclear maser spots
\citep{GG97}.  The masers align along position angle
$\sim -45\arcdeg$, and the high-velocity (red and blue) masers extend
out to $\sim 0.8$~pc (2.7 light-years) symmetrically about the
near-systemic masers. The position-velocity profile of the maser spots
is linear out to a separation of $\sim 0.6$~pc from the near-systemic
masers. The redshifted masers show a falling position-velocity curve
beyond $0.6$~pc; \citet{GG97} claim that the velocities fall more
slowly than $r^{-0.5}$ as would be expected for purely Keplerian
motion. Given the similarities to \objectname[]{NGC~4258}
\citep{Miyoshi95}, the interpretation has been that the
nuclear masers of NGC~1068 also trace a rotating molecular disk
surrounding a central black hole (\citealt{maserpaper};
\citealt{GG97}). 

However, the arguments for NGC~1068 are not as compelling as for
NGC~4258, since the masers do not align at right angles to the radio
jet axis, nor do they clearly align with the thermal disk. One
possibility is that the disk is highly warped between the inner,
ionized region and the outer, molecular region (e.g., as discussed in
the review of the ``NGC 1068 Workshop'' by
\citealt{BB97}). On the other hand, the
nuclear masers might arise from an outflow, such as an AGN-powered
wind (e.g., \citealt{KK94}). In this case, the maser
kinematics are not strictly controlled by gravitation and therefore do
not necessarily measure the central mass. 

\citet{BH96} reported a drift in the velocities of the
red masers over time, which they interpreted as a centripetal
acceleration. This measurement is in conflict with the disk
interpretation. In the disk model, the direction of centripetal
acceleration at the position of the red masers is in the plane of the
sky. There should be no Doppler velocity drifts among the red masers
owing to centripetal motion; rather, there should be detectable
accelerations among only the near-systemic masers. The apparent
accelerations of the red masers therefore argue against the disk
model, but an outflow model may better accommodate their observations.

\subsection{The Jet Masers at Component C}

The jet masers at component C appear to arise near a shock front
between the outflowing radio jet and an intervening molecular cloud
\citep{n1068b}. Apart from the presence of maser emission, further
evidence for this scenario includes the bending of the radio jet axis
by $20\arcdeg$, a localized flattening of the radio spectrum, and a local
brightening of the radio emission. The continuum behavior suggests
either re-acceleration or compression of the synchrotron plasma, or
local magnetic field compression, specifically at this site. An
additional result is that, adopting our suggested radio-HST image
alignment \citep{n1068b}, the brightest scattered optical/UV
continuum occurs at the site of this shock. This result has led us to
propose that the dominant source of scattered light in NGC 1068 is
warm ($T\sim 10^5$~K), highly ionized gas occuring at this shock front
\citep{mythesis}. \citet{Kyoto} and \citet{RCWU98} have since
demonstrated evidence for free-free absorption below 1.6~GHz,
supporting this scenario.

\subsection{The Present Work}

We have re-observed the \h2o\ masers of NGC~1068 with the Very Large 
Array (VLA), and we have monitored the temporal behavior of the maser
emission using the Effelsberg 100m telescope. We present and discuss
the results and analysis of these monitoring studies in this
work. This paper is organized as follows. Section~\ref{obs} describes
the observations, both new and archival, and data reduction
procedures. Sections~\ref{imaging}--\ref{accels} report the results of
the radio spectral line experiments apart from considerations of
reverberation. More specifically, Section~\ref{imaging} reports on the
VLA imaging experiment. Section~\ref{zippo} discusses a VLA search for
high velocity maser emission. In section~\ref{accels}, we analyze the
monitoring data to look for radial accelerations in the nuclear red
masers. The case for reverberation is first discussed in
Section~\ref{reverb}, which includes a study of the correlations of
maser brightnesses and the 1997 flare of the red and blue nuclear
masers. The reverberation results favor a disk geometry, and
Sections~\ref{distance}--\ref{othereffects} discuss the implications of a
disk geometry.  Section~\ref{jet} gives a brief discussion describing
the jet masers associated with radio component C. Finally, we
summarize our main conclusions in Section~\ref{conclusions}.

%% file: obs.tex
\section{Observations and Data Reduction}\label{obs}

The spectral line observations obtained at each telescope and from
archival data employ different conventions for the Doppler velocity
reference frame. Where relevant, we describe below the convention used
for each observation. For the purposes of clarity and comparison, in
the results to follow, we have converted all of the velocities to the
LSR reference frame, using the optical convention for the Doppler
shift.

\subsection{VLA Observations}

We observed the \h2o\ masers of NGC~1068 with the VLA\footnote{The VLA
is operated by the National Radio Astronomy Observatory which is
operated by Associated Universities, Inc., under cooperative agreement
with the National Science Foundation.}  in its A-configuration. The
receivers were tuned to the 22~GHz (1.35~cm wavelength) maser
transition of \h2o\ with appropriate Doppler corrections. The velocity
range spanned by the masers of NGC 1068 is too broad to cover
completely with a single VLA bandpass, so instead we observed
alternating velocities in two 12-hour tracks spanning two days (31
July 1995 and 8 Aug 1995). Specifically, the usable portion of
individual scans cover a velocity range of 148.4~\kms\ with channel
widths of 10.6~\kms. Fourteen such cubes were obtained, with three
channel overlaps between neighboring cubes. The resulting velocity
coverage is 1660~\kms, centered near $cz = 1150$~\kms\ (heliocentric,
optical convention). In addition, we observed NGC 1068 in 22~GHz
continuum tuned well away from the \h2o\ maser emission. The continuum
scans were interleaved with the spectral line scans to ensure similar
$(u,v)$ coverage.

Phase calibration and astrometry were based on intermittent scans of
the nearby calibrator 0237$-$027. We used the 10~mas precision
coordinates from \citet{MPHGA96}.  Excluding statistical and phase
noise, the uncertainty in the absolute astrometric calibration is
therefore $\sim 10$~mas. Amplitude and bandpass calibration were
performed based on scans of the calibrator source 3C~138. This source
is resolved with the VLA-A at 22~GHz, so we performed the amplitude
calibration based on a CLEAN model of a VLA-A observation obtained and
provided by C. P. O'Dea. The flux scaling so bootstrapped is based on
the convention of \citet{BGPW77}. A single iteration of phase-only
self-calibration was applied to the spectral line data, selecting as a
model channels containing bright maser emission, and to the line-free
continuum data set.

The 22~GHz continuum was removed from the spectral line data by
subtracting the Fourier-transform of a CLEAN model of the continuum
data set from the spectral line visibilities (AIPS task
``UVSUB''). Channel maps were then generated using conventional
Fourier-transform techniques and a CLEAN deconvolution. Both
naturally-weighted and uniformly-weighted channel maps were
generated. The rms image noise is $\la 6$~mJy beam\mone\ on uniformly
weighted channel maps and $\la 4.5$~mJy beam\mone\ on naturally
weighted channel maps. These values are close to the predicted thermal
noise levels and represent a factor of roughly 2.5 improvement in
sensitivity over the data presented in \citet{CL86} and
\citet{maserpaper}.

Residual and systematic errors in the primary phase calibration and
self-calibration solutions introduce slight offsets in the relative
astrometry between individual spectral line cubes and the continuum
data. We made channel maps from data sets with and without
self-calibration solutions applied and compared the positions of
individual maser features. The positional differences, measured as an
appropriately weighted rms, are consistent with those expected given
the signal-to-noise, and there is no systematic trend in the offsets
between the data sets with and without self-calibration applied.
Formally, we estimate a conservative (signal-to-noise limited)
uncertainty of $\la 1$~mas in the alignment between the spectral line
data and the continuum data. The uncertainty in the alignment between
overlapping cubes is also $\la 1$~mas based on the positions of
spectral line features detected in the overlapping channels. Judging
from the good alignment of the southern maser features
(Figure~\ref{maserspots}), the cube-cube alignments seem actually to
be much better than these uncertainty estimates imply. For comparison,
the uncertainty in relative positions in the 1983 spectral line data
\citep{maserpaper} is 5~mas between the spectral line cubes and the
continuum image.

\subsection{Effelsberg 100m Spectra}

Using the Max-Planck-Institut f\"ur Radioastronomie (MPIfR) 100m radio
telescope located near Effelsberg, we monitored the 22~GHz \h2o\ maser
spectrum of NGC 1068 over the period 1984--1998. Table~\ref{obstable}
summarizes the observing dates, final spectral coverage, and channel
spacing. Observations prior to 1995 focus mainly on the brighter,
redshifted masers; we have included scans covering the blueshifted
part of the spectrum in the later observations.

We observed scans of bright calibrators every 1--2 hours to monitor
the pointing of the telescope. Typical pointing errors were $\la
10\arcsec$ between pointing checks. The beam-width (HPBW) of the
Effelsberg 100m is roughly 40\arcsec, and therefore the pointing
errors correspond to a signal attenuation of $\la 16\%$.  We used the
18--26~GHz maser receiver, and spectra were generated in a 1024
channel autocorrelator backend.  We usually employed a 50~MHz
bandwidth, resulting in a channel spacing of 0.66~\kms. We commonly
experimented with autocorrelation modes during the course of a run,
either to explore details of narrow emission line features or improve
the signal to noise ratio of a broad emission line feature. 
We ultimately averaged all of the spectra observed in a given
run, irrespective of the variation in channel spacing, to maximize the
signal-to-noise ratio. Spectral averaging required Hanning-smoothing
to the coarsest channel spacing observed in a run and interpolating
the spectra to a common grid using Fast Fourier Transforms. All of the
post-observation processing employed the CLASS data reduction package,
developed jointly by IRAM and the Laboratoire d'Astrophysique de
l'Observatoire de Grenoble (documentation is presently available at
\url{http://iram.fr/doc/doc/doc/class/class.html}). 

To reduce spectral baseline curvature resulting from reflections off
of the secondary support structure, for example, we observed in
position switching mode with a 10\arcmin\ throw. This technique
typically left third order polynomial residuals, which we fit and
subtracted from individual on-off spectral scans, or subscans. Those
subscans showing baselines poorer than third order were discarded
before averaging. The sensitivity of individual subscans varies over
the course of a run, owing mainly to variations in atmospheric opacity
and zenith distance. We therefore chose to subtract baselines prior to
averaging in order to facilitate variance weighting. Variance weighted
averages properly suppress the contribution of noisier subscans and
improve the signal-to-noise ratio of the final, integrated spectrum
relative to an unweighted average. Because each maser spectrum was
covered using two or three separate and overlapping tunings, the
sensitivity varies as a function of velocity across each averaged
spectrum.

We calibrated the flux scale using the standard, 22~GHz gain curve for
the Effelsberg telescope:
\begin{equation}
S_{\nu}({\rm Jy}) = \beta \times \left\{
  \begin{array}{ll}
\left[1 - 0.01\times(30\arcdeg - \theta)\right]^{-1} & \mbox{if $\theta < 30\arcdeg$}\\
1.0                  &\mbox{if $30\arcdeg < \theta < 60\arcdeg$}\\ 
\left[1 - 0.005\times(\theta - 60\arcdeg)\right]^{-1}
    & \mbox{if $\theta > 60\arcdeg$}\\
  \end{array} \right.
\end{equation}
where $\theta$ is the telescope elevation and $\beta = 9.1 (1.0 -
3\times10^{-5} \times v_{LSR})$. This formula applies for Doppler
velocities $< 1.5\times 10^4$~\kms. The frequency dependence of the noise diode
used to establish an initial temperature scale dominates the frequency
dependence of the gain curve.

We estimated the flux calibration uncertainties using two different
techniques. First, we checked the solutions against continuum pointing
scans of flux calibrators including NGC~7027, 3C~48, and 3C~286. The
calibrator fluxes based on the Effelsberg gain curve agree with the
standard values (e.g., \citealt{OWQKSSH94}) to within $20\%$. Second,
we scaled and subtracted spectra from neighboring runs to eliminate
those maser features that did not appear to vary significantly between
runs. The scaling weights were always with the range 85\% -- 115\%,
corresponding to a relative flux scale uncertainty of $\pm
15\%$. Based on this conservative, on-source estimate, we adopt a
formal uncertainty of 15\% for the flux scaling errors.

\subsection{Spectra Published in the Literature}\label{literature}

Going back to 1983, there are several observations of the \h2o\ maser
spectrum of NGC~1068 that are relevant to this study. Using the
freeware utility Windig 2.5\footnote{Windig 2.5 is a Windows
compatible graph digitizer written by Dominique Lovy at the University
of Geneva. It is widely distributed on the internet; we obtained a
copy from \url{http://www.winsite.com}.}, we digitized the published
spectra from scanned bitmaps.  The resolution of the scanned spectra
is degraded slightly relative to the original spectra because of the
finite spatial resolution and sampling of the scanned bitmap
images. The quality of the scanned spectra suffice, however, for the
measurement of spectral peaks or integrated line fluxes. Because of
the limited resolution of the printed and scanned spectra, the line
centroids are not sufficiently accurate to compute radial
accelerations, for example (see \S\ref{accels} for acceleration
analysis of the Effelsberg spectra).

We also digitized the velocity monitoring data of \citet{BH96} and
\citet{NIMMH95}. Converting their data to our adopted Doppler velocity
reference frame (optical -- LSR) required some care. \citet{NIMMH95}
reported their data using the radio convention for Doppler
velocity. \citet{BH96} reported incorrect heliocentric velocities for
NGC~1068. In converting from the LSR frame, they appear to have
subtracted the correction factor of 11.6~\kms, instead of adding. We
converted the velocities appropriately to match our reference frame
and Doppler convention.

References for the published spectra are listed in Table~\ref{pubspectra}. 

\subsection{The Integrated Spectrum}

We averaged all of the spectra from the 1995--1998 Effelsberg
observations to enhance the main spectral features and measure their
recessional velocities. Figure~\ref{redspec} shows the averaged
spectrum for the redshifted masers, and Figure~\ref{bluespec} shows
the spectrum covering the blueshifted nuclear masers, the
near-systemic nuclear masers, and the jet masers.  We measured the
velocities of distinguishable maser features by local Gaussian
decomposition. The results are summarized in Table~\ref{gausstab}.

%% file: results.tex
\section{VLA Imaging of the \h2o\ Masers}\label{imaging}

Figure~\ref{maserspots} plots the VLA, A-array positions of the
nuclear maser spots as a function of channel velocity. Note that these
observations poorly isolate maser features because of the coarse
spatial (0\farcs1) and spectral (10~\kms) resolution of the VLA
data. Nevertheless, we are able to discern the orientation of the
nuclear masers to an accuracy of a few mas. The errors in the maser
positions were estimated as half the beamwidth divided by the
signal-to-noise ratio (e.g.,
\citealt{Fomalont94p219}); this formal error estimate is comparable to
the observed spread among neighboring data points. The orientation in
position angle agrees with the better-suited VLBA observations of
\citet{GG97}, but, since the VLBA \h2o\ maser
observations are self-referenced, and the radio continuum is resolved
out, absolute astrometry of the maser spots is lost in the VLBA
data. The VLA data are phase-referenced against a nearby calibrator
source and therefore retain absolute astrometry accurate to the
uncertainty of the phase calibrator position. In addition, components
NE and C are detected on the VLA channel maps, so we are therefore
able to compare the positions of the masers and the continuum sources.
We find that the position of the individual, systemic masers agrees
with the beam-averaged position of S1 to within 5~mas. Averaging the
positions over the systemic masers, the agreement is better than
3~mas. Unfortunately, radio source S1 is not resolved by the VLA beam
at 22~GHz, and the relative astrometry uncertainty is $\sim 30\%$ the
size of S1 ($\sim 10$~mas) as it appears on VLBA images at 8.4~GHz
\citep{naturepaper}.  We cannot bootstrap a registration between the
VLBA continuum and \h2o\ maser images well enough to discern the projected
geometry. These imaging results are compatible with either a disk or
outflow model for the \h2o\ maser emission, with the near-systemic
masers more nearly centered on the nuclear radio continuum source.

We also confirm the position of the jet masers near the location
of the compact radio component C (cf. \citealt{maserpaper}). The
location of the jet masers is marked on the radio continuum image
in Figure~\ref{continuum}.

\section{VLA Search for High Velocity Masers}\label{zippo}

We expanded the search for 22~GHz \h2o\ maser emission to velocities
$|\Delta v | \la 850$~\kms\ relative to the systemic velocity. The
high-velocity spectrum towards component S1 is included in
Figure~\ref{highv}. We measure no high velocity maser emission to a
sensitivity level of $\sigma \sim 5$~mJy in 10.6~\kms\ channels.
\citet{NIMMH95} have performed an independent search using
the Nobeyama 30m and also failed to detect emission at velocities
between $\Delta v \approx -920$~\kms\ to $+1550$~\kms\ relative to the
systemic velocity; based on their figures, we estimate their
sensitivity was $\sim 10$~mJy in 0.3~\kms\ channels.

For comparison, the \h2o\ spectrum of NGC~4258 shows high velocity, or
``satellite,'' maser features out to $|\Delta v| \ga 1000$~\kms\
\citep{NIM93}. Measured on a linear (non-angular) scale, the maser
distribution of NGC~1068 is $4\times$ larger than the maser disk of
NGC~4258 \citep{GG97}. The lack of high velocity emission beyond $\sim
330$~\kms\ relative to systemic is consistent with the larger size of
the maser distribution if the masers occupy a rotating annulus
surrounding a $1.4\times 10^7$~\Msun\ central mass, roughly 1/3 the
central mass of NGC~4258 \citep{Miyoshi95}.

If X-rays power the nuclear masers, the extent of the maser disk is
determined by the luminosity of the central engine and pressure in the
disk (\citealt{NMC94}; \citealt{MHT96}). Assuming that the
temperatures in the two maser disks are about the same, as is required
by the model to support a large \h2o\ abundance, the pressures in the
maser region scale as $p
\propto (L r^{-2.9})^{1/1.9}$. Based on X-ray observations and models
for electron scattering mirrors, as summarized in \citet{MT97}, the AGN of NGC~1068 is probably about two orders of magnitude
more luminous than that of NGC~4258. Including the result that the
maser distribution is four times larger in NGC~1068, the gas pressure
in the maser disk of NGC~1068 is estimated to be about 40\% greater
than the pressure in the maser disk of NGC~4258. We conclude that the
X-ray heating model is at least compatible with the scaling of the
maser disks in that it implies similar conditions both in NGC~1068
and NGC~4258. 

\section{Search for Radial Accelerations}\label{accels} 

\citet{BH96} found that the redshifted masers appeared
to drift towards higher velocities at a rate of $\sim 1$~\kms\
year\mone. In contrast, a rotating disk model would predict no
measurable acceleration of the redshifted masers, since the direction
of the centripetal acceleration at the kinematic line of nodes is
parallel to the plane of the sky. Radial accelerations should be
present only among the systemic velocity masers, where the centripetal
acceleration vector lies along the line-of-sight. Taking $r =
0.75$~pc and $v_{rot} = 330$~\kms\ \citep{GG97},
the maximum centripetal acceleration should be $\sim 0.2$~\kms\
year\mone, observable only in the near-systemic masers.

Our Effelsberg monitoring spectra span 14 years. If the very high
accelerations measured by \citet{BH96} are correct, we should measure
a dramatic 14~\kms\ drift of the redshifted maser features over this
time period. To measure the centroid velocities of the red maser
features, we decomposed the maser spectra into Gaussians allowing the
amplitude, width, and centroid velocity to vary freely. The properties
of the integrated spectrum were used for the initial
guesses. Figure~\ref{redaccel} plots the fitted centroid velocities
vs. time. The Gaussian fitting technique breaks down in the spectrally
crowded region near 1430~\kms\ where brightness variations of the
overlapping lines affects the centroid fit.  Gaussian fitting
performed well for the bright and relatively isolated 1411~\kms\ line
and other isolated lines, and there is no evidence for radial
acceleration over the course of these observations.

We have included in Figure~\ref{redaccel} the data of \citet{BH96} and 
\citet{NIMMH95}. These data span the gap in our monitoring: the years
1986 -- 1994. \citet{NIMMH95} reported a brief monitoring run in 1992,
and the data do not agree with the contemporaneous observations of
\citet{BH96}. Instead, their velocities agree very well with ours over
the bracketing time periods. The nature of the discrepancy with
\citet{BH96} is not clear. One possibility may be line blending, for
example, between the 1411~\kms\ line and the 1425~\kms\ line. This
possibility does not seem likely, because their average spectrum
appears to resolve that particular blend well. Moreover, line blending
cannot account for the apparent acceleration of the 1453~\kms\
line. It is possible that the 36.6m Haystack antenna was not sensitive
enough to obtain data with sufficient signal-to-noise ratios. Given
the excellent agreement between our data and the data of
\citet{NIMMH95}, only these results were used to calculate velocity drifts.

To measure or constrain the radial acceleration of the red masers, we
performed a least-squares, multi-line fit to the centroid velocity
vs. time data. Maser lines were identified with features in the
integrated spectrum based on proximity in velocity. Transient maser
lines with ambiguous velocity identification were ignored during the
fitting procedure. All fitted lines were constrained to have the same
acceleration. The best fit is consistent with no acceleration, and the
three-sigma upper limit on the magnitude of the acceleration is
0.006~\kms\ year\mone.  We rule out accelerations as large as 1~\kms\
year\mone\ as reported by \citet{BH96}.

The low upper limit on the acceleration of the redshifted masers
favors the rotating disk model and limits the geometry and kinematics
of the masers. The position-velocity curve (Greenhill \& Gwinn 1997)
also favors a rotating annulus geometry. The question is how closely the
high-velocity masers (those on the falling part of the rotation curve)
are located to the line-of-nodes. For the sake of simplicity, if we
assume circular symmetry and an edge-on viewing angle, the apparent
acceleration $a_{obs}$ is related to the true centripetal acceleration
$a_r$ by
\begin{equation}
a_{obs} = a_r \sin{\theta} \,
\end{equation}
where $\theta$ is the angle between the kinematic line-of-nodes and
the radius vector to the location of the red masers within the
annulus. Resolving $a_r = v_r^2 / r$, and accounting for a
$\cos{\theta}$ projection of the radius vector, the constraint on
$\theta$ is then
\begin{eqnarray}
\tan{\theta} & \leq & v_{obs}^{-2} r_{obs} a_{obs} \\
             & \approx & \tan{1\fdg9}
            \left(\frac{v_{obs}}{330{\rm\ km\ s^{-1}}}\right)^{-2}
            \left(\frac{r_{obs}}{0.6{\rm\ pc}}\right)
            \left(\frac{a_{obs}}{0.006{\rm\ km\ s^{-1}\ year^{-1}}}\right)\ ,
\end{eqnarray}
where $v_{obs}$ is the apparent radial velocity and $r_{obs}$ is the
projected radius of the maser features. The normalization of $r_{obs}$
and $v_{obs}$ is based on the disk model described by
\citet{GG97}.  

Unfortunately, there are insufficient data to search for accelerations
of the blue masers; see Figure~\ref{blueaccel}. The prediction remains
that the near-systemic masers should show a drift of $\sim 0.2$~\kms\
yr\mone\ if the rotating disk model is correct. We therefore searched
for accelerations of the near-systemic masers based on our current
monitoring epochs and Effelsberg data taken over 1984--1985.  Among
these earlier observations, we found one 1985 observation that
detected the near-systemic masers. Figure~\ref{sysaccel} plots the
velocity of the near-systemic peak of the 1985 spectrum in reference
to the data from the 1995--1998 campaign. From inspection of this
diagram, looking particularly at the more recent data, it is clear
that the larger velocity changes are not due to centripetal
acceleration. We know from the VLBA observations of \citet{GG97} that
the broad systemic maser line decomposes into a number of narrower
line features and spatially distinct maser spots. It seems that each
of these spots can vary in brightness more or less independently, and,
as a result, the peak of the integrated line profile shifts back and
forth in velocity. In other words, a given near-systemic maser spot
may be brighter than its neighbors for a period of no more than a
few months. Otherwise, lacking observations between 1986 and 1984,
there is no discernable trend in the near-systemic velocities that
would indicate acceleration. On the other hand, we also cannot rule
out acceleration of the near-systemic masers with the present data.

\section{Evidence for Reverberation}\label{reverb}

\subsection{Motivation}

\citet{CL86} pointed out that the entire
maser spectrum of NGC~1068 faded between 1983 and 1984, and they
originally argued that the masers were responding to a common maser
pump source whose efficiency was also fading during that time. Neufeld
et al. (1994) have since argued that X-rays produced by the central
AGN might provide the energy for a maser pump (see also
\citealt{WW97}). X-ray heating produces an extended region of warm dust
at a few hundred degrees K. These temperatures favor gas phase H$_2$O
production, and the H$_2$O abundance is raised. The H$_2$O molecules
receive pump energy from excited H$_2$ molecules by
collision. Variations of the maser luminosities might therefore
result, in part, from variations of the central X-ray source, akin to
the reverberation of broad optical line emission in active galaxies
(e.g., \citealt{NP97p85}). This scenario might be complicated,
however, by local effects, such as shock heating or chance cloud
alignments (e.g., \citealt{WW94};
\citealt{MM98}). Such local effects might be unrelated to variations
of the central pump source, effectively adding noise to any
reverberation signal.

In analyzing VLA observations of the \h2o\ masers of NGC~1068
\citep{maserpaper} and the new VLA observations presented here, we
noticed that the trend originally suggested by \citet{CL86} continued
between 1983 and 1995; that is, the blue, near-systemic, and red
masers appeared to brighten and dim coherently as measured on time
scales of $\ga 2$~years. The dynamical time between the blue and red
masers is $\sim 10^4$~years, but the light travel time is only $\sim
2.5$ years between the central engine and the maser spots. It is
plausible that we are witnessing variations of the maser brightnesses
in reverberation to variations of the central, obscured AGN.

This reverberation model makes simple predictions depending on the
geometry of the maser sources.  If, as is commonly thought and
supported here, the geometry is a uniformly rotating disk, the
near-systemic masers should trace gas along our sight-line to the AGN,
and the red and blue masers should trace warm molecular gas along the
forward quadrants of the disk and along the line-of-nodes, or disk
``midline;'' see Figure~\ref{toyannulus} for illustration. The reasons
for placing the high velocity masers along the disk midline are
because (1) the projected velocity profile falls with radius for these
masers, (2) the longest velocity-coherent optical paths occur at the
disk midline, and (3) we measure no significant (projected) radial
acceleration of these masers, implying that the high velocity masers
lie within $\sim 2\arcdeg$ of the kinematic line of nodes
(Section~\ref{accels}).  The high velocity masers along the line of
nodes are roughly 0.6~pc more distant than the near-systemic
masers. Variations of the blue and red masers should be nearly
simultaneous but should lag behind varations of the near-systemic
masers by $\ga 1.9$~years.

For a bipolar outflow geometry, the prediction is that the blue and
near-systemic masers should vary nearly simultaneously, but the red
masers should lag behind. The time lag for the outflow geometry is not
predictable since the orientation of the outflow relative to the
sight-line is unknown. If the outflow is symmetric about the central
source, the time lag would measure the orientation.  The absence of
any obvious lags on $\sim 2$~year time scales then argues that the angle
between the polar axis and the line-of-sight should be greater than
$\ga 45\arcdeg$ (i.e., more nearly in the plane of the sky).

Evidence for reverberation would support the idea that the AGN powers
the \h2o\ masers, perhaps, as suggested above, indirectly via X-ray
heating \citep{NMC94}. Another conceivable power source might be
AGN-driven explosions and subsequent blast waves that heat the
molecular medium through shocks (e.g., \citealt{EHM89}). Reverberation
imaging offers a means of measuring the geometry of the nuclear \h2o\
maser source and a geometric estimate of the distance to
NGC~1068. Testing the reverberation model is challenging because the
nuclear masers spread over a distance of several light-years. Compared
with BLR reverberation, therefore, time lags within the \h2o\ maser
spectrum may be observationally expensive to track, requiring years of
regular monitoring. It may also be difficult to distinguish locally
generated maser flares from reverberation events.

\subsection{Analysis}

One important concern for reverberation analysis is the uniformity of
the flux calibrations among the monitoring spectra. For example, it
may be that the masers appear to vary coherently because systematic
errors in the flux scale artificially brighten or dim all of the maser
features simultaneously. Accurate flux calibration is challenging at
22~GHz owing to variations in atmospheric opacity and pointing errors,
and, since the original goals of the observations were different from
the goals of this study, no additional effort was made to ensure a
stable flux scale.

We cannot improve upon the ($\sim 15\%$) accuracy of the relative flux
scale for which errors. We can instead use the jet masers as a control
to check whether errors in the flux scale might give rise to the
apparent coherence of the nuclear maser variations. The jet masers are
located $\ga 30$~pc away from the red, blue, and near-systemic masers
associated with the nucleus. Even if the jet masers were powered by
the same pump as the nuclear masers, the time lag would be on the
order of a century. It seems more likely, however, that the jet masers
are powered independently by shocks
\citep{maserpaper}. Since the data span
only $\sim 15$~years, we yet lack sufficient information to address
coherence between the jet and nuclear masers. Any apparent
coherence between the jet masers and the nuclear masers would
therefore result from errors in the flux scale. 

Unfortunately there are yet insufficient data to perform a 
cross-correlation analysis as is traditionally applied to reverberation
data. Instead, we explored the temporal coherence of the nuclear
masers both by (1) qualitatively comparing the time variations of the
nuclear and jet masers and (2) evaluating the correlation of the
measured peak fluxes and luminosities of the masers. For each of the
correlation tests we chose the red nuclear masers as the reference
because they are the brightest masers and because they were always
observed throughout the monitoring program.

The reason for distinguishing between the peaks and total luminosities
in the correlation tests is because the profiles of the red, systemic,
and blue maser emission profiles decompose into many narrow, spatially
distinct emission features. We do not know {\em a priori} how the
timescale for variability of the maser pump source, $\tau_{pump}$,
compares to the timescale for propagation of the pump signal through
the maser clouds, $\tau_{prop}$. If $\tau_{pump} < \tau_{prop}$, then
variations in the maser response will be radially localized, and
reverberation will appear more clearly in the variation of the maser
line peaks. On the other hand, if $\tau_{pump} > \tau_{prop}$,
variations of the maser response will be more nearly uniform across
the maser clouds, and evidence for reverberation may also appear in
the variation of the integrated luminosities.

Looking first at the time variations, Figure~\ref{lumtime} compares
the variability of the luminosities of the nuclear masers and the jet
masers as a function of time.  Figure~\ref{fluxtime} plots the
flux densities of the peak maser emission as a function of time. In
both figures the luminosities and fluxes were normalized to their mean
over the observing period better to compare with the variations of the
red masers, which are a factor of ten or so brighter than the other
maser groups. In the comparison of the maser line peaks, the maser
spectra were smoothed to the lowest common resolution of all of the
observations, or 10~\kms\ (the spectral resolution of the VLA data).

The maser luminosities vary by factors of three or so over the
monitoring period. It is clear that all of the masers, including the
jet masers, vary more widely than would be allowed for systematic
errors in the flux calibration. There appear flux variations at least
as rapid as the shortest time sampling, or roughly two weeks.
Figures~\ref{lumtime} and \ref{fluxtime} qualitatively demonstrate
that the jet masers vary independently of the nuclear masers. Note
that the reported luminosities were computed assuming isotropic
emission. In particular, the jet masers show jumps in luminosity over
the observing period 1992--1996 that are not in proportion to the light
curves of the nuclear masers. Looking at the broader trend, the peak
of the jet maser emission dims over 1990 -- 1998; this trend
is not present in the light curves of the nuclear maser peaks. There
appears to be no strong bias introduced by errors in the flux
calibration.

Next, to assess the coherence of the nuclear maser variability more
quantitatively, we measured correlations between the near-systemic,
blue, and jet maser fluxes with the red maser fluxes. The
reverberation model predicts that the peak fluxes and luminosities of
the nuclear masers should correlate (i.e., the blue masers are
brighter when the red masers are brighter), but the jet masers should
not correlate with the nuclear masers.  Figure~\ref{lumlum} compares
the integrated luminosities, and Figure~\ref{fluxflux} compares the
peak maser flux densities. (Note that the reported luminosities were
computed assuming isotropic emission.) To test for correlation, we
computed Pearson's linear correlation coefficient, Spearman's rank
correlation coefficient, Cox's proportional hazard model, and
Kendall's generalized $\tau$ to measure the correlations between these
flux measurements. The latter two statistics properly account for the
two upper limits (e.g.,
\citealt{IFN86}), but these upper limits were ignored in deriving the
former two statistics. Table~\ref{corrtab} lists the correlation
statistics for the peak-peak and luminosity-luminosity comparisons.

Table~\ref{corrtab} reports two significance levels. The first is the formal
probability that the measured correlations could have occurred by
chance (i.e., a low probability argues in favor of a real
correlation).  As a check, we also performed Monte Carlo simulations to
estimate the significance level. The Monte Carlo simulations involved
generating artificial observations by randomly varying the fluxes or
luminosities assuming a Gaussian distribution of errors. To remove any
true correlations in the data, we also scrambled the order of the
simulated observations relative to each other. For example, the blue
masers would have been randomly reordered relative to the red masers
to remove the apparent correlation between their fluxes and
luminosities. We then repeated the computation of the statistics for
$10^4$ simulated observations and counted the fraction whose
correlation coefficients met or exceeded the those measured for the
real observations. The large number of Monte Carlo trials was intended
mainly to sample the measurement errors. Note that the so-derived
probabilities of chance correlation are overestimated because an
occasional trial may not have sufficiently removed real correlations
in the data. Nevertheless, the Monte Carlo significance levels compare
well with the formal significance levels, and therefore the
correlation statistics are robust even though the sample numbers are
small and the measurement errors are occasionally large.

We measure significant (better than $3\sigma$) correlation only
between the blue and red nuclear masers, although the Monte Carlo
statistics for the Pearson $r$ statistic give a $\sim 4\%$ probability
of no correlation. The significance of correlation between the
near-systemic and red masers is only $\sim 2\sigma$ -- $3\sigma$. As
we would predict based on the qualitative comparison, the jet masers
show no correlation with the red nuclear masers in any test, and we
conclude that systematic errors in the flux scale have not caused the
measured correlations between the nuclear maser groups. 

The nuclear masers show coherent variations on timescales shorter than
the length of the monitoring period (Figure~\ref{fluxtime}). We also
conclude, therefore, that the measured correlations do not result
solely from undersampling of long-term trends, i.e., variability
trends longer than the monitoring period. That we measure trends both
in the peak fluxes and total luminosities of the nuclear masers
suggests that $\tau_{pump} \ga \tau_{prop}$, or, that the pump signal
propagates through clouds faster than the variation time-scale of the
central pump source, as argued above. The light travel time through
the maser annulus is $\sim 1.9$~years, and so it appears that the time
scale for variations of the central pump source is typically longer.

In summary, the measured correlations are in accord with the
predictions of the reverberation model. These results cannot, however,
distinguish between the disk and outflow models, since the time
sampling was usually coarser than 2~years, the predicted
light-diameter in the disk model, throughout the monitoring period. We
next consider a single, short flare event of the red and blue nuclear
masers that favors the disk model.

\subsection{The 1997 \h2o\ Maser Flare}

Over 3 -- 4 Nov. 1997, we detected a flare of $\sim 300$~mJy in the
blue masers at 859~\kms. For comparison, the average peak flux density
of the blue masers in recent measurements has been $\sim
30$~mJy. Figure~\ref{maserflare} plots the history of this flare
between Oct. 1997 and Feb. 1998. There also appears a corresponding
flare of the 1411~\kms\ maser, although it is more difficult to see
because of confusion with neighboring maser emission. We enhanced the
flare spectra by taking difference spectra between successive
observing epochs. Figure~\ref{maserflare2} plots the difference
spectra between Oct. 1997 and Feb. 1998. The central (mean) velocity of the
flares is $v_{LSR} = 1134$~\kms\ (corresponding to a heliocentric
velocity $v_{\sun} = 1145$~\kms), consistent with the systemic
velocity of the host galaxy based on \hi\ imaging studies
\citep{BSTT97}. Another flare appears at 1376~\kms\ in the 3 Jan. 1998
spectra, but there is no clear connection between this event and the
Nov. 1997 flares.


The reverberation model predicts simultaneous red and blue flares if
the geometry is a rotating disk, but blue flares should precede red
flares in an outflow geometry. Alternatively, the observed red and
blue flares may be responses to independent flares of the pump
source. For example, the red flare may be responding to an earlier
flare of the pump source, and the blue flare may be responding to a
later flare of the pump source.  An outflow geometry would require
that the pump source flares are separated by a time interval that
matches the light travel time along the line-of-sight between the red
and blue maser sources. Since the pump source should have no knowledge
of maser clouds' orientation relative to our sight-line, the observed
simultaneous flares would have to be coincidental, but we cannot rule
out this possibility on the strength of a single flare event. The
rotating disk model remains, nevertheless, the simpler explanation,
because the simultaneity of the flares is explained only by geometry
and requires no coincidental timing of the maser pump and light-travel
delay. We conclude that the data favor the rotating disk model for the
\h2o\ maser disk geometry.

%% file: discussion.tex
\section{The Predicted Behavior of Nuclear Maser Flares: Towards a
Geometric Distance Estimate}\label{distance} 

Based on the evidence and arguments presented above, our favored model
for the variations of the nuclear masers is a rotating annulus
responding in reverberation to a common, central pump source, probably
the central engine. Figure~\ref{toyannulus} depicts the model geometry
for the nuclear masers based on the VLBA observations of
\citet{GG97}. We next consider some implications and predictions based
on this model and models for the pump source. For brevity and the
purpose of the discussion, we assume the disk model to be correct.

Looking first at the VLBA imaging results of \citet{GG97}, the masers
are distributed more or less in a straight line on the sky, running
blue--systemic--red from southeast--northwest
(cf. Figures~\ref{maserspots} and \ref{toyannulus}). There is a gap
between the blue and systemic masers, but faint maser emission fills
the region between the brighter systemic and red maser emission. In
projection, the Doppler velocities rise linearly with angular
separation between the systemic and red masers. The velocity curve
turns over beyond $\sim 8.6$~mas from the systemic maser peak. This
region of the velocity curve is compatible with the kinematics of a
rotating annulus, in which the linear part of the velocity curve
traces masers at a (nearly) common radius. The falling part of the
velocity curve traces masers distributed along the disk line-of-nodes,
i.e., nearly parallel with the sky plane and spanning increasing
radii. We refer to masers on the rising part of the rotation curve as
the ``edge masers,'' and masers on the falling part are called the
``midline masers.''

The implication is that the edge masers span nearly a quarter of the
rotating annulus (note that Figure~\ref{toyannulus} plots only the
brightest maser features from the integrated spectrum; see
\citealt{GG97} for a plot of all of the maser spots). We assume these
masers trace the forward (i.e., nearer) quadrant, since free-free
absorption may attenuate \h2o\ maser emission on the far side (e.g.,
\citealt{KL89}; Neufeld et al. 1994; Gallimore et al. 1997). One
prediction of the reverberation model, therefore, is that maser flares
should appear to accelerate, starting in the systemic masers and
proceeding through the peak of the red masers, as we receive the
response of the edge masers at increasing distances. This apparent
acceleration is due solely to time-lag geometry, and the mean
apparent acceleration should be very rapid, $\sim 140$~\kms\
year\mone. Assuming an edge-on disk and carrying out the time-lag
geometry, the velocity of flares along the inner edge of the disk as a
function of time-delay is given by:
\begin{equation}
v(t) = v_0 \pm v_{rot} \times \left[ \frac{2ct}{R} \left( 1 -
\frac{ct}{2R} \right) \right]^{1/2}\ ,
\end{equation}
where $v_0$ is the systemic velocity, $v_{rot}$ is the rotation
velocity of the annulus, and $R$ is the radius of the annulus. A sum
is taken for the redshifted masers, and a difference is taken for the
blueshifted masers. The lag time is normalized to zero for the
response of the near-systemic masers. 

After this apparent acceleration event, there should be an apparent
deceleration event as the midline masers respond in succession. After
the edge masers have flared, the time delay is no longer due to the
increasing distance of the masers from us, since the midline masers
are assumed all to be at roughly the same distance. Instead, the time
delay is due to the increasing distances of the midline masers from
the central engine.  Again assuming an edge-on disk, and further
accounting for the propagation of signal along the line-of-nodes, the
response of maser emission along the line-of-nodes should follow:
\begin{equation}
v(t) = v_0 \pm v_{rot} \times \left[ \eta \left(
\frac{ct}{R} - 1 \right) + 1  \right]^{\alpha},
\end{equation}\label{eq-lon}
where all variables are defined above, except $\eta$, which is the
ratio of the propagation speed of the pump signal to the speed of
light, and $\alpha$, which is the power law index for the
position-velocity curve: $v \propto r^{\alpha}$; $\alpha = -0.5$ for a
Keplerian disk. 

Figure~\ref{drifts} plots the flare velocity as a function of
time-delay, or velocity-delays curve, appropriate for NGC~1068 based
on the analysis above. In constructing this plot, we assumed $\eta =
1$ and $\alpha = -0.5$, but these quantities remain to be measured. 

\subsection{Deriving a Geometric Distance from a Single Flare Event}

The velocity-delays curve ($v[\Delta t]$) is sensitive to the assumed
distance to NGC~1068. To illustrate, the dotted lines in
Figure~\ref{drifts} trace the velocity-delays curve for distances of
12.7~Mpc and 18~Mpc. The delays curve effectively measures the
physical radius of the annulus, and VLBI measurements provide the
angular radius. Tracing the full velocity-delays curve for a single
flare event may therefore provide a geometric distance measure to
NGC~1068.  The main uncertainty, as with all such techniques, would be
the geometry of the maser sources. For example, the assumption of
circular symmetry also factors into the geometric distance estimate to
the \h2o\ masers of NGC~4258 \citep{HMGDINMHR99}. 

Using reverberation to measure the distance to NGC~1068 offers an
advantage over the use of centripetal accelerations or proper motions
because, firstly, the systemic masers are faint, requiring long
integration times to measure velocity drifts or proper
motion. Secondly, the maser disk of NGC~1068 is large compared to the
disk of NGC~4258, and so the centripetal accelerations and proper
motions are predicted to be correspondingly smaller than found in
NGC~4258.

Once a flare of the near-systemic masers is discovered, it would
suffice to observe the spectrum at $\sim 1$ month intervals of $\sim
2$~years to trace out the velocity-delays curve accurately. One caveat
is that the difference spectra might be contaminated by flares
unrelated to the reverberation mechanism; such flares might result
from, e.g., inhomogeneities in the maser clouds, chance cloud
alignments, or local excitation effects such as shocks. In addition, a
careful measurement would initially require frequent observations
to improve estimates of the predicted delays and durations of
the reverberating flares. The procedure would be iterative, with
initial observations providing an estimate of the distance, from which
one could predict more precisely the delay to the next flare, and so
on, ultimately to reduce the observational expense of the later
observations. 

The November 1997 flare occurred among the midline masers at 859~\kms\
and 1411~\kms, and therefore we may find later reverberations among
the midline masers over the following 90 days (assuming a distance of
14.4 Mpc). Specifically, we would have expected flares at 1408~\kms\
around 2 December 1997; 1404~\kms\ around 5 January 1998; and
1395~\kms\ around 27 January 1998.  Our only near-overlap is the 3
January 1998 observation that should have detected a flare of the
1404~\kms\ maser. That we did not detect the flare does not rule out
the reverberation model; it may be that the flare duration was shorter
than a day, or, more likely, that the distance to NGC~1068 is not
exactly 14.4~Mpc. Although there appear weaker flares at 1376~\kms\ on
3 Jan 1998 and 1425~\kms\ on 31 Jan 1998 (Figure~\ref{maserflare2}),
they fall outside the reverberation time window and are probably
unrelated to the simultaneous flare event of 3 Nov 1997. 


\section{The Symmetry of the Maser Disk}

Returning to the 1997 \h2o\ maser flare, it is somewhat surprising
that the blue and red flares appeared simultaneously. We would have
expected some asymmetry to the disk resulting in a corresponding lag
between the maser flares. We can use the simultaneity of the flares
and limits on the flare duration to constrain the symmetry of the
disk. No flares appeared on the Oct. 12 spectrum, 22 days before the
flare detection, and no flares appeared on the Jan. 3 spectrum, 60
days following the the flare detection. The maximum duration for the
flare is therefore 82 days. Supposing, without loss of generality,
that the red masers were fading just as the blue masers were
brightening, and further assuming that the red and blue masers flared
for the same duration, the maximum delay between the red and blue
flares is 22 days. The fractional deviation in the disk radius to the
blue and red masers is the ratio of the lag between the maser flares
($< 22$ days) to the travel time of the pump signal. 

If X-ray heating provides the pump energy, the signal travel time to the
1411~\kms\ maser is $\sim 2.6$~years. Between the 1411 and 859~\kms\
masers, the maser disk would deviate no more than $2.3\%$. Factoring
in the uncertainty in the systemic velocity, we find that the radial
deviation is $\la 4\%$. Note that this is a robust upper limit,
because a slower pump signal, such as a blast wave might provide,
would require a greater degree of symmetry. For example, a 1000~\kms\
blast wave would require radial deviations $\la 0.01\%$ (not
accounting for the uncertainty in the systemic velocity). It is hard
to imagine a parsec-scale disk to be so symmetric; we favor the X-ray
heating model mainly because it places the most relaxed limits on the
symmetry of the disk.

\section{The Case for Dust Reverberation}\label{ircraziness} 

Optically thick, hot dust at $T_d \sim 1000$~K, probably located near
the nuclear maser region (e.g., \citealt{NM95}), is thought to be a
significant source of near-infrared continuum emission from the
nucleus of NGC~1068 (e.g., \citealt{PK93}). Supporting this picture,
the upper limit on the size of the nuclear near-infrared source is
$\sim 1.4$~pc, based on speckle imaging with the Keck telescope
\citep{WNM99}. This upper limit is comparable to the size of the \h2o\
maser emission and radio continuum emission thought to arise from the
warm region of the obscuring disk (Gallimore et al. 1997). Since the
dust and masers appear to be nearly co-spatial, then, if the \h2o\
masers disappeared as the result of a flare of the bolometric
luminosity of the central engine, the prediction is that near-infrared
emission should show a nearly contemporaneous brightening as the dust
temperature rises in response to the same flare \citep{Neufeld00}. 

\subsection{Variability of the Near-Infrared Continuum}

\citet{Glass97} has monitored the variability of NGC~1068 in
near-infrared bands over a period spanning the \h2o\ monitoring
experiment of \citet{BH96} and the new Effelsberg observations
presented here. Figure~\ref{irplot} compares the variability of the
K-band (2.2\micron) continuum emission with the variability of the 1411~\kms\
midline maser. Relevant to this line of argument,
\citet{Glass97} originally suggested that the near-infrared
light-curve might follow the response of near-nuclear dust to a steady
increase in the UV--X-ray luminosity of the central engine.  For
comparison, evidence for the response, or reverberation, of
near-nuclear dust to variations in the UV luminosity of the central
engine has been observed in other Seyfert galaxies, such as Fairall 9
(\citealt{CWG89}; \citealt{Barvainis92}) and Mrk~744
(\citealt{Nelson96}), among others. In contrast, we cannot see the
central engine of NGC~1068 directly, and so we cannot directly compare
the light-curves of the near-IR emission and maser emission with the
light-curve of the central engine. The challenge, then, is to find a
model that can reproduce the maser and near-IR variability and make
predictions for other observable effects resulting from the
variability of the hidden central engine.

Examining Figure~\ref{irplot}, the 2.2\micron\ luminosity of NGC~1068
has been increasing steadily since 1976, peaking around 1994--1995,
and showing a decline from 1995--1998. Looking at the broader trend,
the 1411~\kms\ maser dims until its disappearance in 1990
and brightens after its reappearance in 1991. The only possibility to
consider is whether the peak of the infrared light curve corresponds
to the loss of maser signal in 1990. The implied time-lag between
observed events is $\sim 4$~years.

The predicted time lag between the response of the masers and the
near-IR continuum depends on the orientation of the obscuring disk and
optical depth effects. For example, if the obscuring disk is optically
thin at near-infrared wavelengths, then an initial peak of the
near-infrared continuum should have preceded the loss of midline maser
signal by $\sim 3$~years. There is no evidence for a peak of the
2.2\micron\ light-curve over that time period (around 1987), and so we
rule out this possibility. Instead, the interesting case to consider
is whether the disk is optically thick to near-IR radiation, so that the
near side of the hot dust region is obscured by cooler dust in the
foreground; see Figure~\ref{irschematic} for illustration. There would
be an unattenuated sight-line only to the far edge of the maser disk,
and the 2.2\micron\ peak should lag behind the loss of maser
signal. In this optically-thick model, the lag time will also depend
on the inner radius of the hot dust region, which can itself depend on
variations of the central X-ray luminosity.

\subsection{A Model for the Near-Infrared Variability} \label{irmodel}

To test the plausibility of this scenario, we constructed a numerical
model to calculate the observed near-IR light curve from an inclined,
optically thick torus. We assumed that the inner edge (or surface) of
the disk was fixed by the sublimation temperature of dust grains
(similar to the ``reformation'' models used by \citealt{Barvainis92}
and \citealt{Nelson96}). As the X-ray luminosity of the central engine
rises, the local dust temperature rises. Grains are heated above the
sublimation temperature, rapidly evaporate, and the inner edge of the
disk moves outward. Because grains do not exceed the sublimation
temperature, the emissivity of the inner edge of the disk remains
constant. However, the total near-IR luminosity rises because, as the
inner edge of the disk moves outwards, the total exposed surface area
of the far side of the disk increases. This model also requires that
grain reformation is rapid after the ionizing flux wanes in order to
restore the conditions necessary to generate masers. For reference,
the conditions of \h2o\ maser clouds ($n\sim 10^9$~cm\mthree\ and
$T\sim $ few hundred K; e.g., \citealt{RM88}; Neufeld et al. 1994)
compares to the conditions under which grains form in classical nova
shells (e.g., \citealt{NH78}). The timescale for grain formation in
novae is $\sim 2$ -- 3 months (e.g., \citealt{Bode88} and references
therein), which is short relative to the slow variability of the
near-infrared light curve of NGC~1068.

We computed the time lag of emission from a region of the disk according to
\begin{equation}
c \Delta t = r_{disk} + z\ ,
\end{equation}
where $r_{disk}$ is the separation between the central engine and the
disk region, $z$ is the distance of the disk region from us, and $c$
is the speed of light. We did not account a delay 
for grain evaporation, because the evaporation time-scale is much
shorter than the observed near-IR variability \citep{Voit91}; in other
words, the response of the inner edge of the disk to changes in the
luminosity of the central engine was taken to be immediate. Based on
the observed separation between the masers and the central engine, the
disappearance of the red masers over 1990 -- 1991 predicts that the
peak of the central engine light curve should have occurred around the
decimal calendar date 1987.9. The time of the peak of the flare was
therefore fixed to that date.  We assumed a gaussian profile for the
temporal shape of the central flare. We also adopted the nuclear
spectrum derived by \citet{spectrum} and, for simplicity, assumed that
the spectral shape was preserved during the flare.

For the purposes of this model, the geometry of the obscuring disk was
taken to be a cylindrical annulus with uniform gas density. The gas
composition is probably clumpy, and the masers probably reside in
particularly dense clouds within the dusty medium. The densities that
go into the model are therefore effectively averages weighted by
emissivity and opacity. We used the code DUSTY \citep{NIE99} to
estimate the dust temperature as a function of opacity into the model
disk. To account for different sublimation temperatures, we ran a grid
of models varying the dust temperature of the inner edge of the
annular disk. The distribution of dust grains was taken to be the
standard, galactic distribution \citep{MRN77}, which is approximated
within DUSTY by single-sized grains with suitably weighted average
properties. In principle, the inner edge of the disk should have a
stratified composition owing to the dependence of the sublimation
temperature on grain composition \citep{Salpeter77}, but accounting
for the stratification is computationally expensive. Instead, we
assumed a uniform composition and, in preliminary models, fit the
observed near-IR data for the effective dust temperature at the inner
edge. We found that $T_{sub} \sim 1300$~K produced the best fits to
the data, and we fixed this value for subsequent runs of the fitting
program.  This effective sublimation temperature is reasonable given
that composition dependent sublimation temperatures are estimated to
be in the range 1000 -- 2000~K \citep{Salpeter77}.

To compute the model infrared light-curve, our code numerically
integrated the radiative transfer equation along sight-lines through
the inclined model disk, again properly accounting for the light
travel-time from the central engine, to the disk region, and then to
the observer. The integrated near-IR flux was computed at each
observation date of \citet{Glass97} and compared to the measured
near-IR flux. Starlight contributes significantly to the J-band
(1.2\micron) and H-band (1.6\micron) fluxes, and colder dust
contributes to the L-band (3.3\micron) fluxes. Because the
contribution of starlight and cold dust to these bands is not
currently known, we chose only to fit the K-band (2.2\micron) fluxes.

Based on their speckle observations, Weinberger et al. (1999)
decomposed the nuclear K-band emission into three components:
starlight, which makes up about 600~mJy within a 12\arcsec\ aperture;
a nuclear point source, presumably the nuclear disk, which at the time
of their observations (1995) made up about 275~mJy; and an
extended source, oriented roughly in the direction of the radio jet,
with a flux density of about 275~mJy. Their 12\arcsec\ aperture flux
agrees well with the measurements of \citet{Glass97} observed near the
same epoch (Fig~\ref{irplot}). However, both the infrared light-curve
and HST/NICMOS observations may be at odds with this speckle imaging
result. Looking first at the light curve, in 1976 the
starlight-subtracted K-band flux density was about 180~mJy, and the
peak in 1994 was $\sim 700$~mJy\citep{Glass97}. Weinberger et
al. (1999) made their measurements near the peak of the near-infrared
light curves, and so the nuclear point source can account for no more
than $\sim 275$~mJy of variability. The remaining 245~mJy worth of
variability would have to arise from the extended K-band source. The
size of the extended source is however $\sim 75$~LY, and it would be
impossible to reconcile such variability over a period of 18~years.
To examine further the reality of the extended 2.2\micron\ source, we
are currently analyzing archival HST/NICMOS images of the nucleus of
NGC~1068. Within the central 3\arcsec, the NICMOS data are compatible
with a smooth stellar background and a nuclear source smaller than
0\farcs1. Any additional, extended flux must be weaker than $\sim
15\%$ of the point source flux (Gallimore \& Matthews, in
preparation).

Accepting that the detailed structure of the near-IR source is
uncertain, we instead modeled the K-band light-curve as comprising two
components: 600~mJy from starlight (or other extended, non-varying
background) and the remainder from a parsec-scale nuclear disk. We fit
our model light curves to Glass' data using an iterative simplex
algorithm to minimize $\chi^2$ (e.g.,
\citealt{numericalrecipes}). The results are displayed in
Figure~\ref{irresponse}, and the best fit parameters for the model are
listed in Table~\ref{t_irmodel}. Accepting that the model
oversimplifies the geometry of the disk and the shape of the X-ray
flare, the fit is good: $\chi^2 = 50$ with 20 degrees of freedom, and
the model reasonably reproduces the shape of the light curve. 

We estimated the statistical uncertainties of the fit by exploring the
$\chi^2$ values as a function of the parameters taken at intermediate
iterations of the simplex algorithm. Because this method only samples
the $\chi^2$ values traced by the converging simplex, the statistical
uncertainties reported in Table~\ref{t_irmodel} are biased but
probably of the right order of magnitude. The systematic errors are,
of course, much greater than the statistical uncertainties. We
repeated the fitting procedure, taking into account the following
sources of systematic error: (1) choice of integration algorithm for
the radiative transfer equation, (2) the response of the simplex to
initial guesses for the parameters; (3) the assumed shape of the flare
profile; (4) the assumed contribution of starlight; (5) the assumed
disk scale height; and (6) the spatial resolution of the integration
grid. For variations of the flare profile, we considered three
additional profile shapes: tophat, sawtooth, and sawtooth plus step
(the latter allowing for an overall change in the luminosity after the
flare). The corresponding variations are reported as systematic errors
in Table~\ref{t_irmodel}. Note that we have by no means exhausted
uncertainties owing to model assumptions: other sources of systematic
error include, for example, the assumption of constant density and
strict cylindrical geometry. Our integration algorithm depends on
shortcuts derived from these assumptions, however, and so it was not
practical to test these sources of error.

To fit both (1) the four-year lag between the loss of maser signal and
the peak of the near-IR light-curve and (2) the change in the infrared
luminosity, we find that the central engine flared for $\sim 10$~years
duration (FWHM), peaking at nearly five times the initial
luminosity. These parameters result in a plausible and self-consistent
model connecting a (hidden) flare of the bolometric luminosity of the
central engine, the loss of maser signal in 1990 (based on the model
of \citealt{Neufeld00}), and the near-IR light curve, which peaks in
late 1994. However, the monitoring data span a period of time that is
not much greater than the $\sim 10$~year duration of the flare. Based
on the strength of only a single maser -- near-IR event, we cannot
rule out coincidence. It may take decades of additional monitoring to
find another near-IR reverberation event as a test of this model.

\section{Other Observable Effects of the Reverberation
Model}\label{othereffects} 

Looking for other effects of a flare of the X-ray or bolometric
luminosity, one might also expect a brightening of the observable
cooling lines (e.g., \boiiib) of the narrow line region (NLR). Based
on the radio-optical registration of MERLIN and Hubble Space Telescope
images (e.g., \citealt{n1068b}; \citealt{CML97};
\citealt{Kishimoto99}), the central ten parsecs of the NLR appear to
be obscured by foreground dust. As a result, the response of the NLR
to the postulated X-ray flare may not yet be apparent, and it will be
difficult to disentangle the delays. 

Exploring the variations of the obscured central engine will also be
difficult. The electron scattering ``mirror'' is located, in projection,
$\sim 20$~pc away from the location of the nuclear masers. Although
the response may be somewhat diluted by the extent of the scattering
medium, the prediction is a time-lagged flare in the directly
detectable X-ray emission, perhaps some 60~years hence.

\section{Comments on the Origin of the Jet Masers}\label{jet}

Although the jet masers are not the focus of the present work, we
digress to note that they are also variable over the period of this
monitoring experiment. The variability is not so remarkable, in the
sense that maser emission associated with molecular outflows from
young stellar objects (YSOs) may vary by two orders of magnitude, or
disappear entirely, on timescales of months (e.g.,
\citealt{CWBWMT96}). The comparison stops there, because the
outflow velocity of the jet of NGC~1068 is probably $\ga 10^4$~\kms\
\citep{n1068b}, whereas the outflow velocities of YSOs probably do not
exceed a few tens of \kms\ (e.g., \citealt{RSBC99}). Models for \h2o\
maser generation, whether by X-ray heating or shocks, generally
require the survival of catalytic dust grains (e.g., \citealt{EHM89};
\citealt{NMC94}). Gas temperatures may elevate to a few hundred~K at which the
gas phase production of \h2o\ is enhanced. The concern for jet-induced
masers is that dust grains do not survive shocks faster than $\sim
100$~\kms\ (e.g., \citealt{JTHM94}).

An alternative explanation for the variability is that the jet masers
amplify variable radio continuum emission from component~C . This
model does not require that the molecular gas is physically associated
with the radio jet. One strike against this model is that there is no
evidence for variability of the radio continuum emission from
component~C, based on \citet{n1068a} and our own, unpublished MERLIN,
VLA, and VLBA observations. Our limit on the variability of
component~C is $\la 5\%$. For comparison, the jet masers vary by
roughly a factor of two. Therefore, the jet masers would have to be
amplifying the radio continuum by a factor $> 40$ to accomodate the
small continuum variations. 

We can use this amplification limit to constrain the covering fraction
$f_C$ of maser clouds against component C. The typical peak maser flux
density is $\sim 65$~mJy, and these masers amplify a continuum signal
$f_C \times 54$~mJy. Adopting an amplification of $G > 40$, then $f_C
< 0.03$.  Owing to the low variability of the continuum source, the
background amplification model would require compact maser
clouds. We cannot rule out this model based on the present data, but
we do not favor it on the grounds that it does not explain how the
masers are pumped, or why they should occur coincidentally in front of
a bright continuum source offset from the nucleus.

We propose an alternative scenario in which the masers occur in a
warm, molecular precursor region heated by radiation from the
shock. For example, \citet{TW86} modeled \h2o\ maser
emission from molecular precursors in star forming regions. They
argued that dust grains might be heated by infrared radiation from the
shock, leading to, at least temporarily, conditions of hot dust mixed
with cold gas. Infrared emission from the dust grains leads to
radiative pumping of \h2o, thereby fueling maser emission
\citep{GK74}. The difficulty with this model, however,
is that the IR photon budget of NGC~1068 may not be sufficient to
power the observed megamasers \citep{CL86}. Instead, we suggest that
the molecular precursor region is heated by X-radiation emitted by hot
post-shock gas. For the jet outflow velocities estimated for NGC~1068
and other Seyfert galaxies, the predicted temperatures of post-shock
gas are in the 1--10~keV range (cf. \citealt{DS95b}). The X-ray
emission from the shock might power maser emission in the surrounding
molecular gas by the process outlined by Neufeld et al. (1994). The
critical parameters are the density and velocity of the jet, which
determine the X-ray luminosity, and the pressure in the surrounding
medium. Unfortunately, the jet properties are poorly constrained, and
so it is difficult to assess this scenario quantitatively. Better
treatment will require hydrodynamical simulations, and numerical
studies might be inverted to constrain the conditions in the jet. Such
a simulation is outside the bounds of the present work.

\section{Conclusions}~\label{conclusions}

We have located the positions of the \h2o\ masers of NGC 1068 relative
to the 22~GHz continuum emission with an overall accuracy of $\sim
5$~mas, comparable to the sizes of the VLBI continuum components, but
much smaller than the ($\sim 300$~mas) separation between adjacent
VLBI components. Most of the maser emission lies within about 10~mas
of the nuclear continuum feature S1 (the nuclear masers), and there is
also maser emission associated with the jet continuum feature C (the
jet masers).

We measure no significant (projected) radial accelerations of the
nuclear masers; the upper limit on the magnitude of the observed
acceleration is $0.006$~\kms\ year\mone. If we assume a rotating disk
geometry, this limit places the high velocity masers within 1\fdg9 of
the kinematic line of nodes.

The brightnesses of the nuclear masers are correlated on time-scales
of years, and they are uncorrelated with variations of the jet maser
brightness. The time-scale for variations of the nuclear masers can be
at least as rapid as a few weeks. Coherent variations occur on time-scales
shorter than the dynamical time but comparable to the light crossing radius,
roughly 2 years, of the source. These results argue in favor of
a reverberation model, in which the nuclear masers are responding to a
common, central pump source. 

We find that there is better correlation between the red and blue
maser brightnesses than the red and systemic maser brightnesses. We
also detected a single flare event in the red and blue masers, but
there is no corresponding flare in the systemic masers. These results
are compatible with a rotating annulus geometry for the nuclear
masers.

We also find that the correlations between the nuclear masers are
noisy. It is clear that local effects must contribute to the maser
variability. Such local effects probably include disk structure,
shocks resulting from cloud-cloud collisions, and projected
cloud-cloud alignments. Nevertheless, that there are measurable
correlations in the peak fluxes and luminosities of the nuclear masers
argues that a central pump source plays a dominant role. In addition,
since the spatially-integrated luminosities are correlated, the
time-scale for propagation of the pump signal through the maser
annulus, $\tau_{prop} \ga 1.4 $~years, should be shorter than
the variation time-scale of the pump, $\tau_{pump}$.

We have constructed a model explaining the four-year lag between the
loss of maser signal in 1990--1991 and the peak of the near-IR K-band
light-curve in 1994--1995. The model requires that the disk is highly
inclined and opaque to near-IR radiation, so that we can only see
emission from hot ($\sim 1000$~K) dust grains on the far side of the inner
surface of the disk. In this model, the central engine brightens over
the course of about ten years, peaking in late 1987. The signal
reaches the red masers about three years later, raising the dust
temperature of the maser region, which should reduce the luminosity of
the \h2o\ masers \citep{Neufeld00}. The signal also causes the inner
radius of the disk to expand as dust grains evaporate. It takes
another four years for the far side of the disk's inner edge to swell
to its maximum size, at which point we see observe the peak of the
near-IR light curve. This model cannot account for the extended,
non-stellar flux in the observations of Weinberger et al. (1999),
which anyway may be at odds with the observed K-band light curve and
HST/NICMOS observation. Although this model provides a plausible
connection between the \h2o\ and near-IR variability, we cannot rule
out coincidence given the observation of a single candidate event.

For future work, a primary test of the reverberation model would be to
detect lags between the near-systemic masers and the blueshifted and
redshifted masers. Detections of time lags between the nuclear masers
would also measure the geometry of the disk perpendicular to the
sky-plane, but as argued above, this may be observationally
challenging owing to the size of the disk. On the other hand,
tracing the velocity-delays curve for one such flare may provide a
geometric measure of the distance to NGC~1068.

Determining the nature of the pump source may also prove
challenging. To test the X-ray pumping model of Neufeld et al. (1994),
for example, would require nearly simultaneous monitoring of the
nuclear X-ray emission and the \h2o\ masers. However, it appears that
the circumnuclear disk of NGC~1068 is completely Compton-opaque even
to hard X-rays.  Nuclear X-ray emission is instead observed in
scattered light (e.g., \citealt{spectrum} and references
therein). Because the size of the scattering region extends several
hundred parsecs (e.g., \citealt{MGM91}; \citealt{Capetti95b}), any
flares in the X-ray spectrum may be diluted by the the range of
time-lags spanned by the scattering medium.  Monitoring radio
continuum flares is similarly difficult, because the circumnuclear
disk is opaque by free-free absorption longward of $\sim 6$~cm
wavelength. In fact, based on our own, unpublished VLA and MERLIN
observations monitoring 6 and 1.3~cm wavelengths, there appears to be
no variability of the nuclear radio continuum components over the past
two decades. Any further tests of the reverberation model will
probably depend mainly on time-lags in the \h2o\ maser spectrum and
corresponding echos in the near-infrared variability.

We also considered possible origins for the jet masers. It seems
likely that they are related to a shock front between the radio jet
and molecular gas. However, since dust would probably not survive the
shock, H$_2$ does not easily reform in the cooling, postshock gas, and
it is not clear how \h2o\ masers would be pumped. We suggest that the
masers might instead arise in an X-ray heated precursor
region. Further tests of this model will probably rely on numerical
simulations of jet shocks in molecular gas.

%% file: ack.tex
\acknowledgements{ We would like to acknowledge the following
colleagues for helpful discussions, correspondence, and comments: Eric Agol,
Julian Krolik, Ari Laor, Lynn Matthews, David Neufeld, Phil Maloney, Barry
Turner, Niranjan Thatte, and Bill Watson. We also thank the anonymous referee
for a careful reading of the manuscript and advice that greatly improved the
presentation and discussion. We are grateful to Lincoln Greenhill, who
helped us determine the sky location of bright maser features detected
in the single dish spectra. J.F.G. received support from a Jansky
Postdoctoral Fellowship, awarded by the NRAO, and by Bucknell
University. J.F.G. dedicates this work to the memory of Dr. Paul Meyer. }

%% file: fig.tex
\begin{figure}\singlespace
\plotone{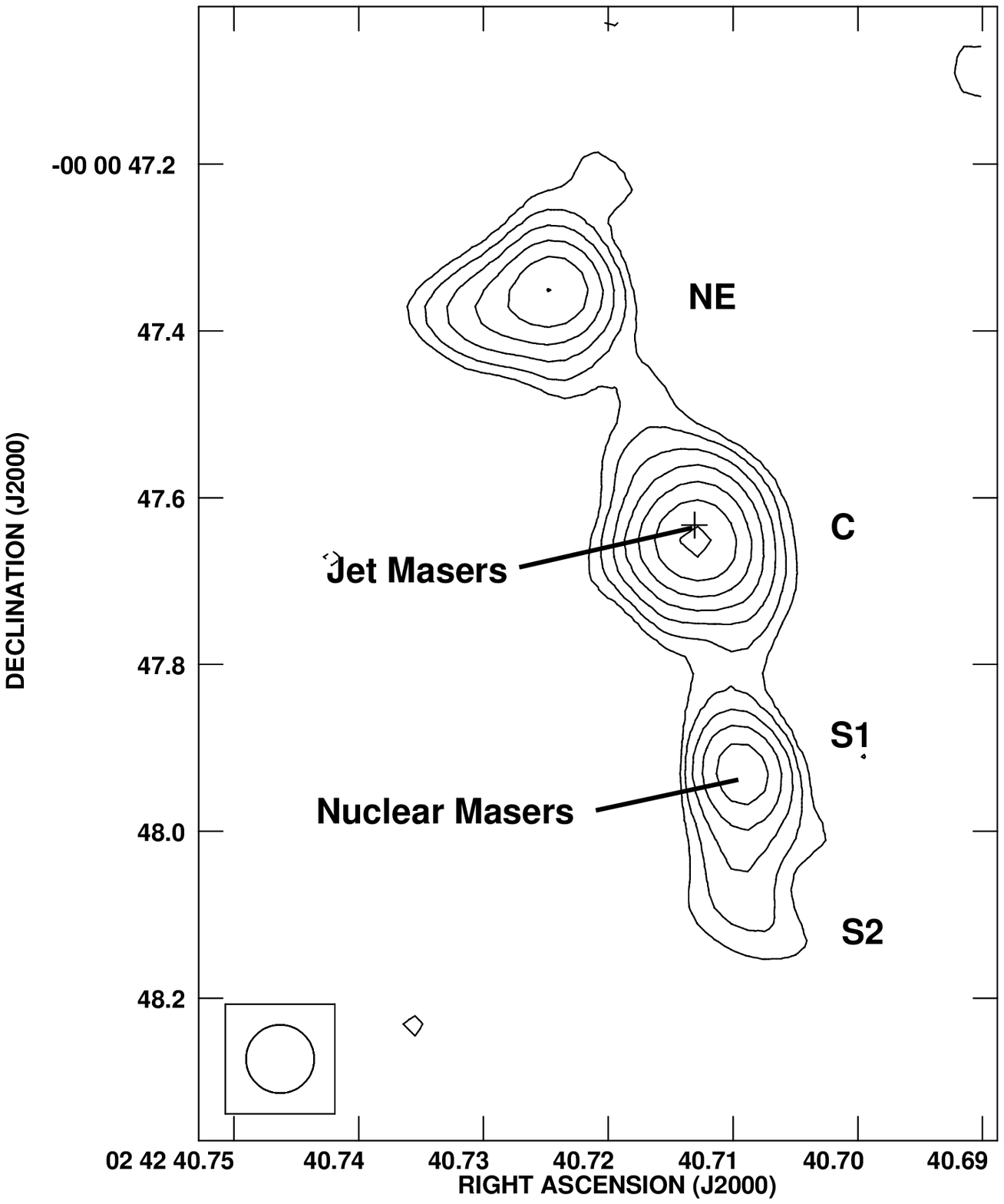}
\caption{New VLA-A 22~GHz continuum image of the subarcsecond jet of
NGC~1068. The contour levels are: $-1.7$, $1.7$ ($\sim 3\sigma$),
2.9, 4.9, 8.2, 13.9, 23.4, \& 39.5~mJy. The restoring beam size
is 82~mas (FWHM), circular.}
\label{continuum}
\end{figure}

\begin{figure}\singlespace
\plotone{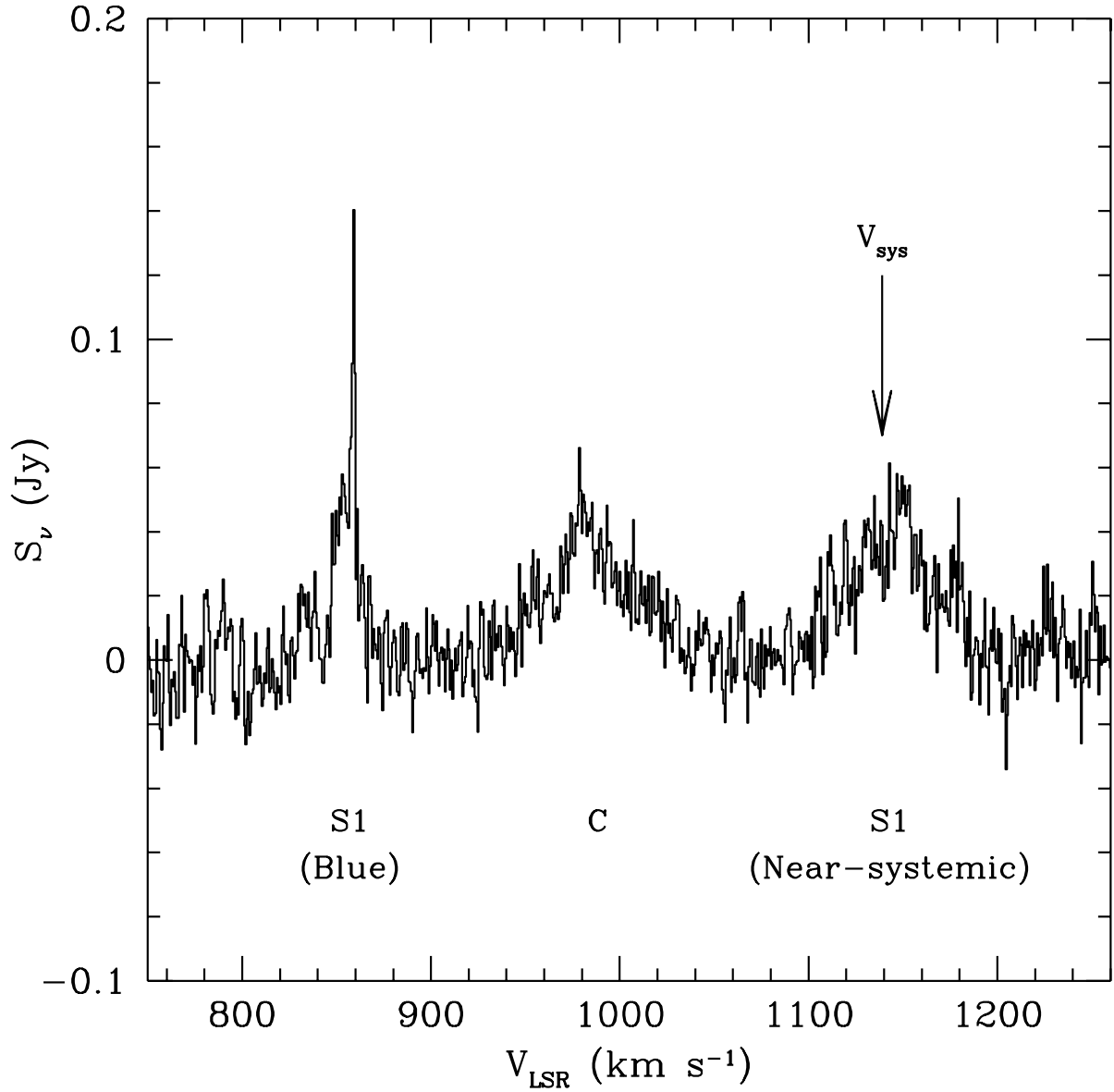}
\caption{Integrated \h2o\ maser spectrum of NGC~1068 spanning the blue
and near-systemic nuclear masers and the jet masers at component C.  This
spectrum was derived from a variance weighted average of Effelsberg 100m
observations spanning 1995--1998. The names of the associated radio
continuum components are labeled below the spectrum (see
\protect\citealt{maserpaper}). }
\label{bluespec}
\end{figure}

\begin{figure}\singlespace
\plotone{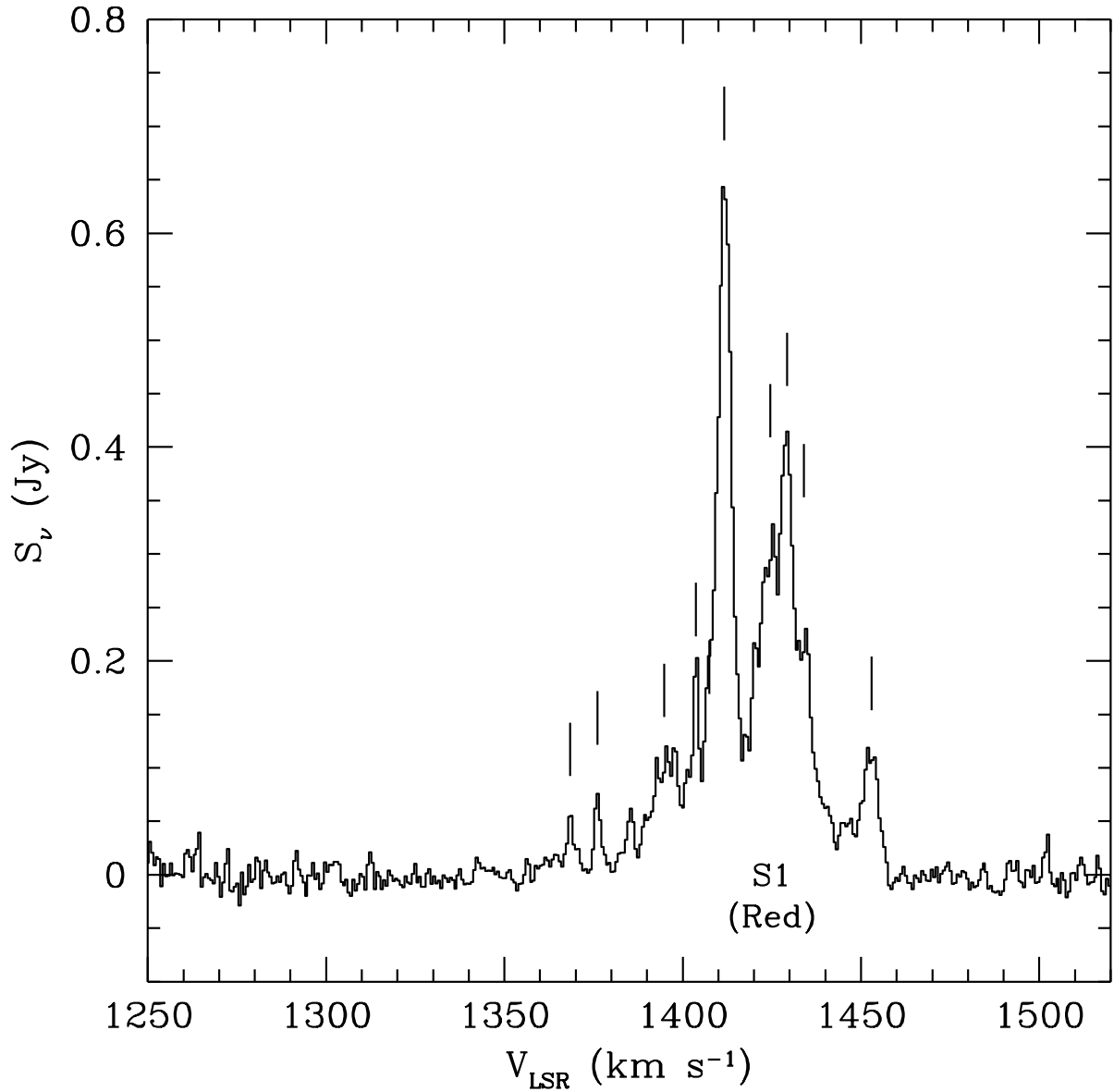}
\caption{Integrated \h2o\ maser spectrum of NGC~1068. Only the
red nuclear masers are presented in this figure. This spectrum
was derived from a variance weighted average of Effelsberg 100m
observations spanning 1995--1998. The plotting convention is as in
Figure~\protect\ref{bluespec}, except that the velocities of
distinguishable maser features, listed in Table~\ref{gausstab}, are
indicated by vertical bars.} 
\label{redspec}
\end{figure}

\begin{figure}\singlespace
\plotone{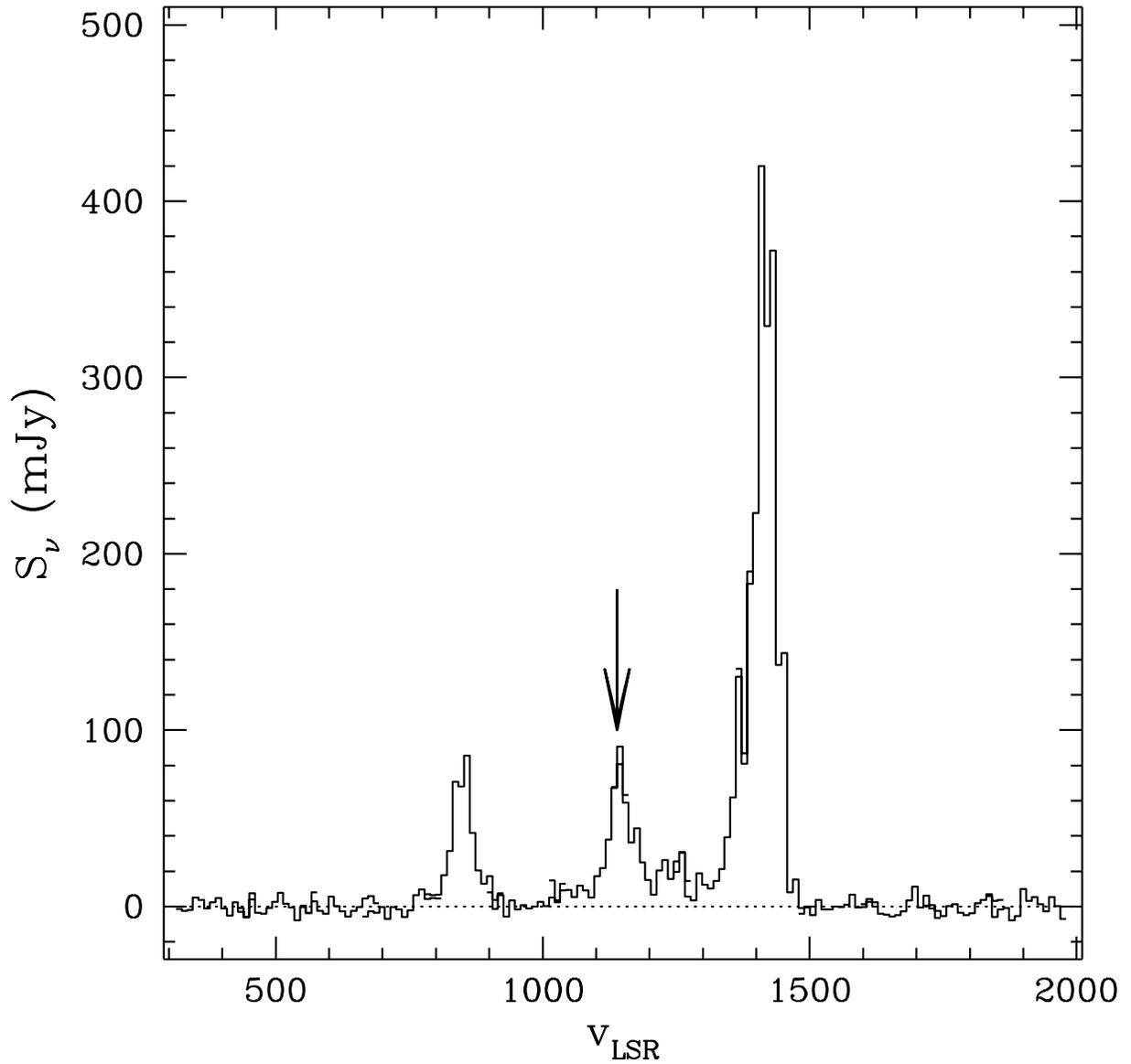}
\caption{The 1995 VLA spectrum of the nuclear radio component S1. The
vertical arrow marks the systemic velocity of the host galaxy. The
observations comprise 14 overlapping tunings. Each tuning is plotted,
including the overlap regions. The discrepancies between the
overlapping spectra are within the measured noise. }
\label{highv}
\end{figure}

\begin{figure}\singlespace
\plotone{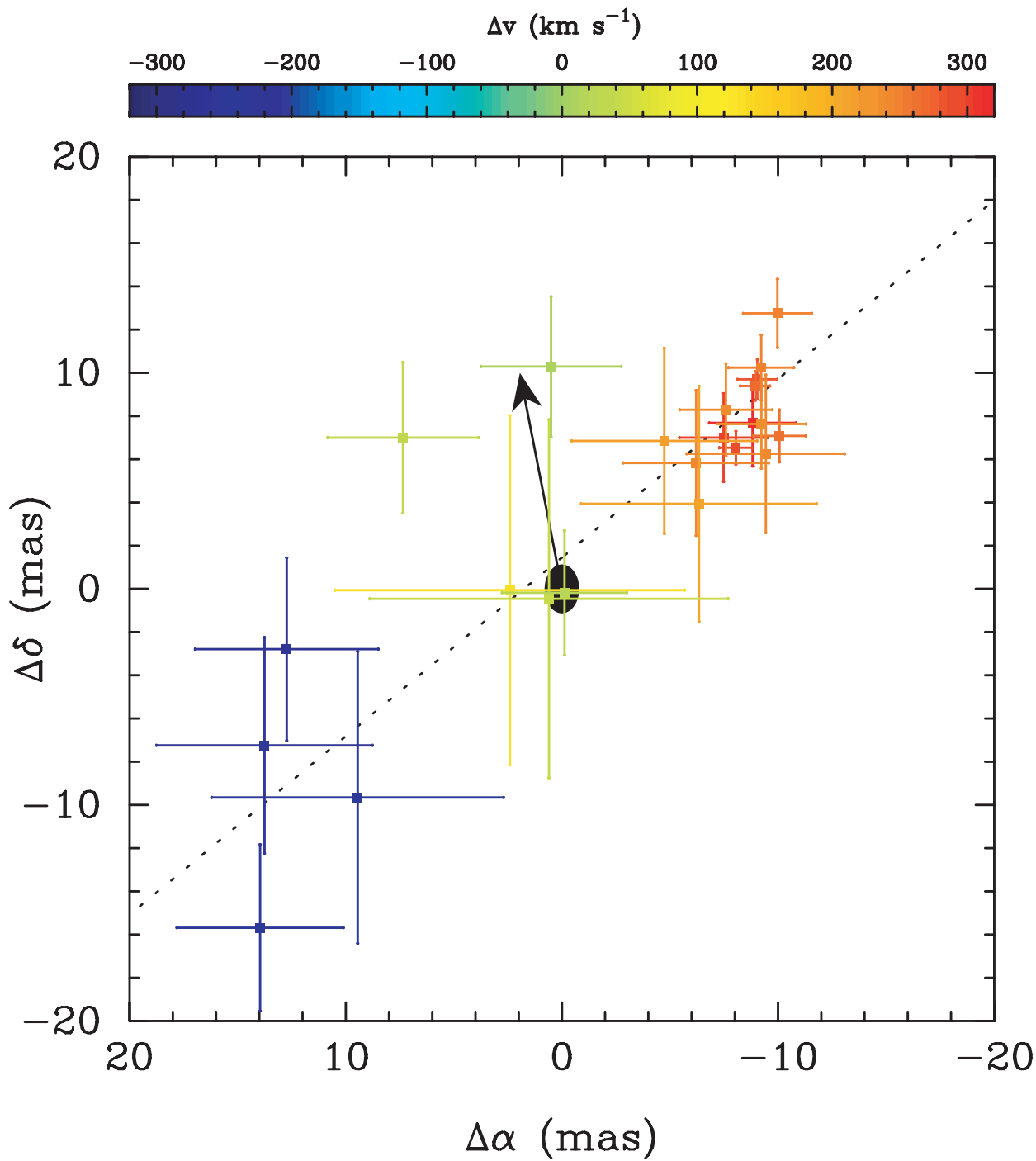}
\caption{The positions of the southern maser spots relative to the
local continuum peak S1. The position of S1 is indicated by the solid
black $1\sigma$ error ellipse with an arrow indicating the local jet
axis towards component C. Note that S1 probably extends $\pm 10$~mas,
based on VLBA imaging at 8.4~GHz (Gallimore et al. 1997).
The maser spots are color-coded according to
velocity ranges relative to $v_{LSR}= 1139$~\kms\ and are marked with
$1\sigma$ errorbars. The errorbars are based on the signal-to-noise
and do not include the relative astrometic uncertainty, $\la 1$~mas,
between successive spectral line cubes and the radio continuum
image. The long dashed line is the best fit maser axis
(P.A.~$-40\deg$).}
\label{maserspots}
\end{figure}

\begin{figure}\singlespace
\plotone{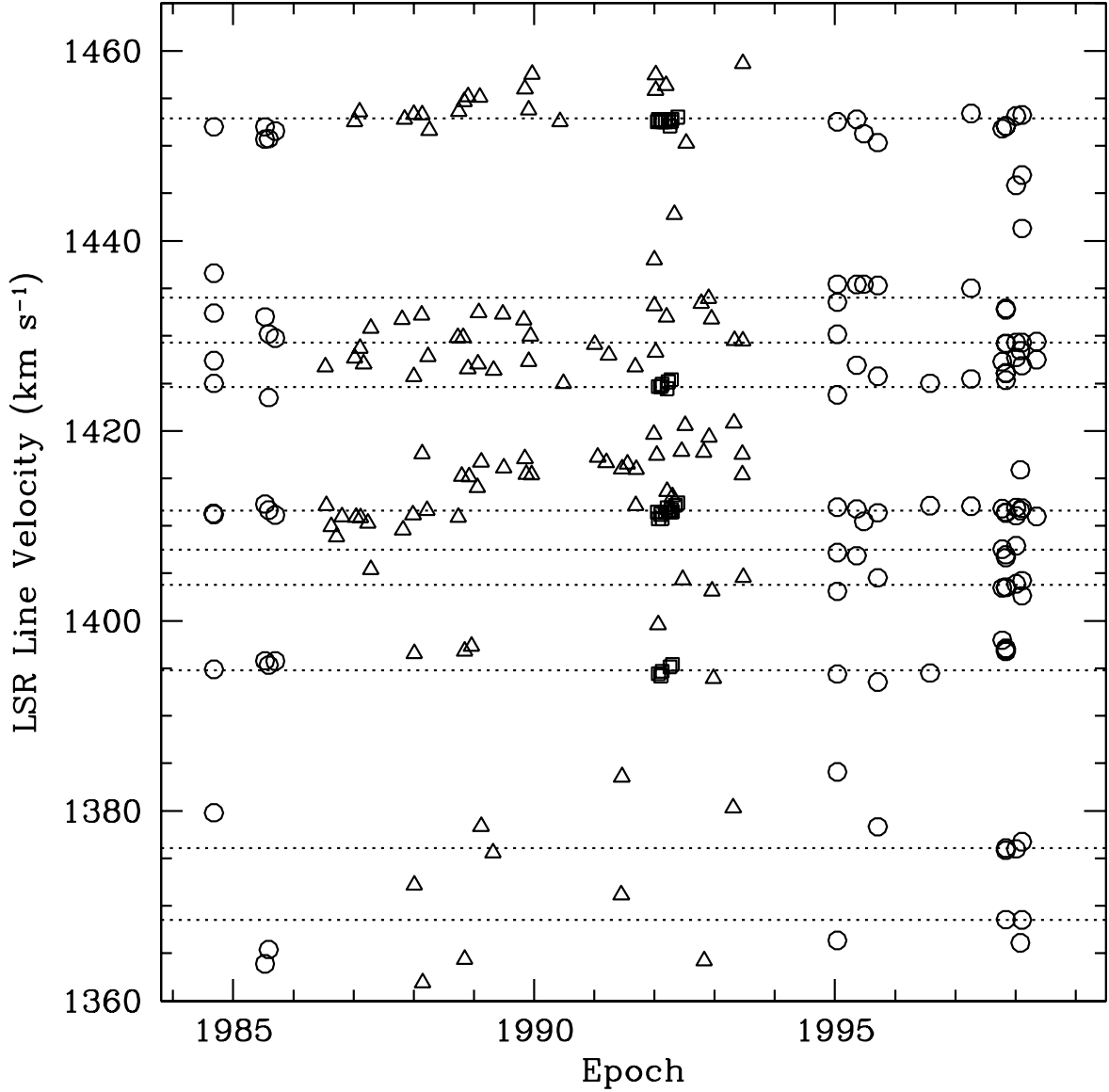}
\caption{Radial accelerations of the red nuclear masers.  A plot of
the centroid velocities of distinguishable maser lines as a function
of observing epoch. The symbols represent the different data sources:
circles mark our monitoring data, triangles mark the monitoring data
of \citet{BH96}, and the (tightly crowded) squares mark the monitoring
data of \citet{NIMMH95} (see Section \ref{literature} for an
explanation of the velocity corrections). The formal uncertainty of
our centroid velocity measurements is typically $\la 0.1$~\kms. The
dotted lines mark maser velocities from the integrated Effelsberg
spectrum. We measure a $3\sigma$ upper limit of $|a| < 0.13$~\kms\
year\ for the radial accelerations of the red masers. 
}
\label{redaccel}
\end{figure}

\begin{figure}\singlespace
\plotone{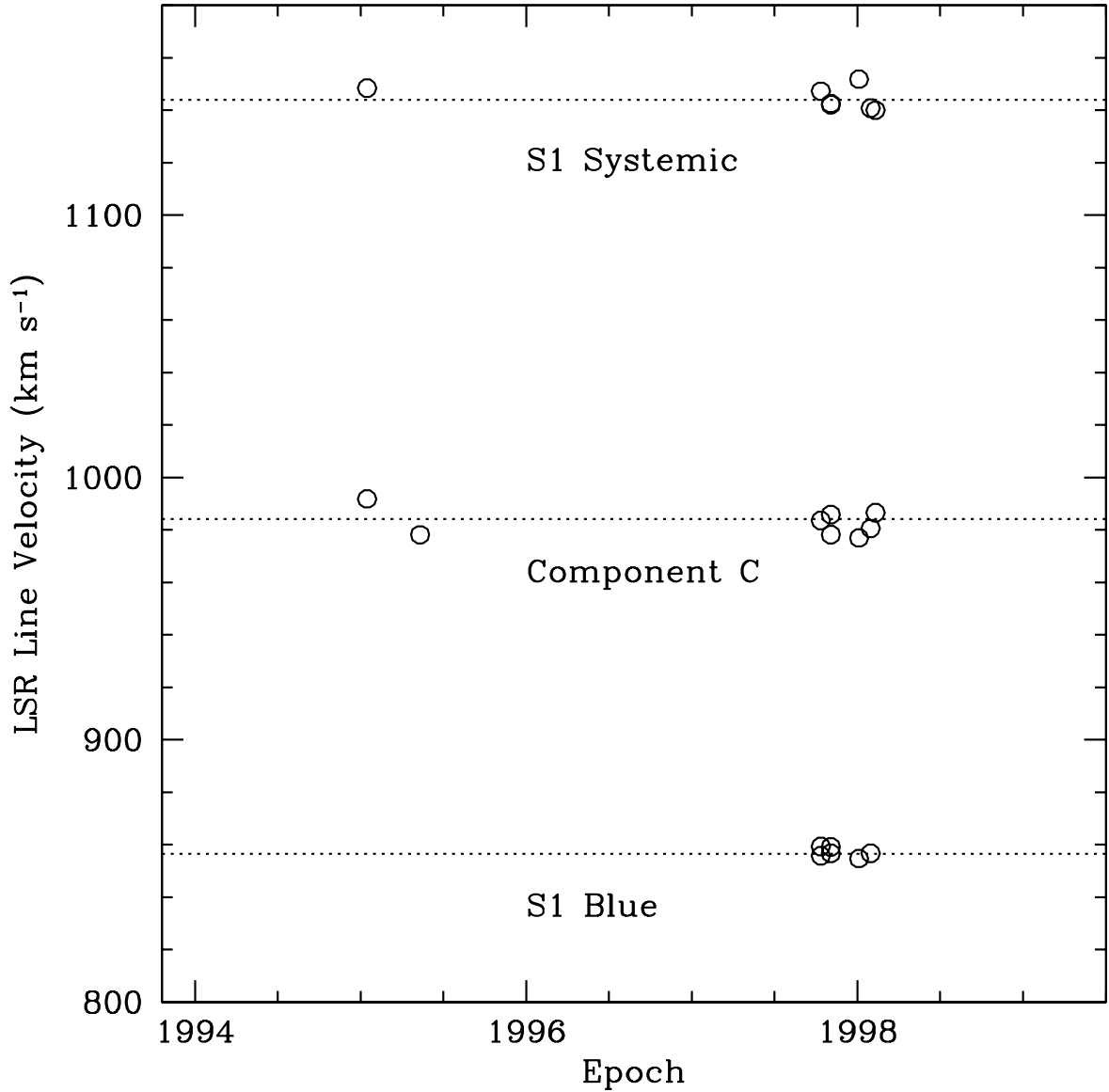}
\caption{Radial accelerations of the blue and systemic nuclear and the
jet masers (component C). The plotting convention is that of
Figure~\protect{\ref{redaccel}}. Currently, there are insufficient data
for a meaningful measurement of the accelerations for these groups of
masers.  The dotted lines mark maser velocities from the Effelsberg
integrated spectrum.
}
\label{blueaccel}
\end{figure}

\begin{figure}\singlespace
\plotone{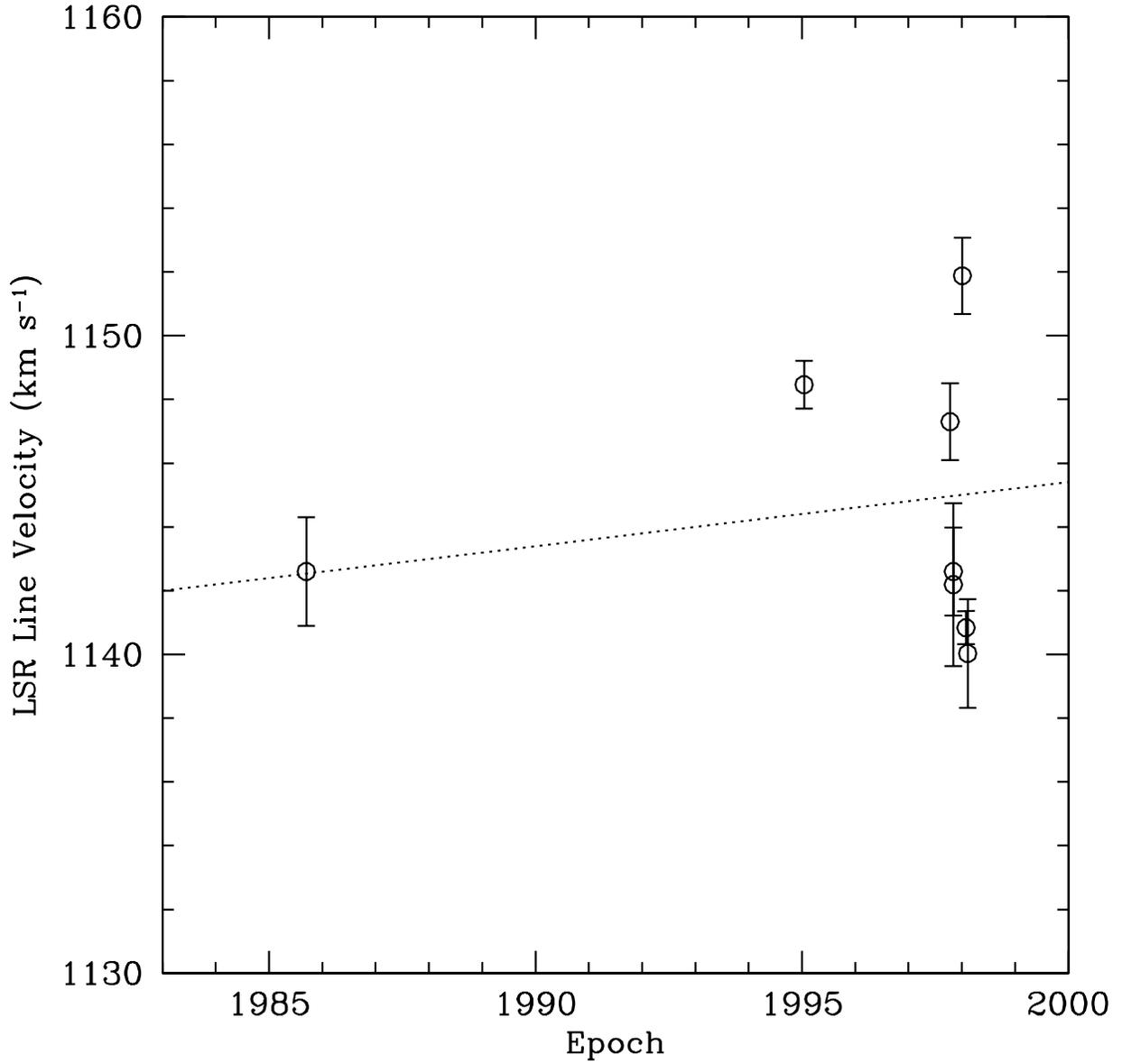}
\caption{Radial accelerations of the near-systemic nuclear masers including a
measurement taken in 1985. The dotted line traces the predicted
acceleration assuming a distance of 14.4 Mpc and circular symmetry of
the maser disk. Based mainly on the distribution of velocities between
1995 and 1998, it appears that the changes in velocity owe mainly to
random variations of the brightnesses of individual maser
spots. Lacking data for 1986--1994, we cannot find a trend ascribable
to centripetal accelerations. 
}
\label{sysaccel}
\end{figure}

\begin{figure}\singlespace
\plotone{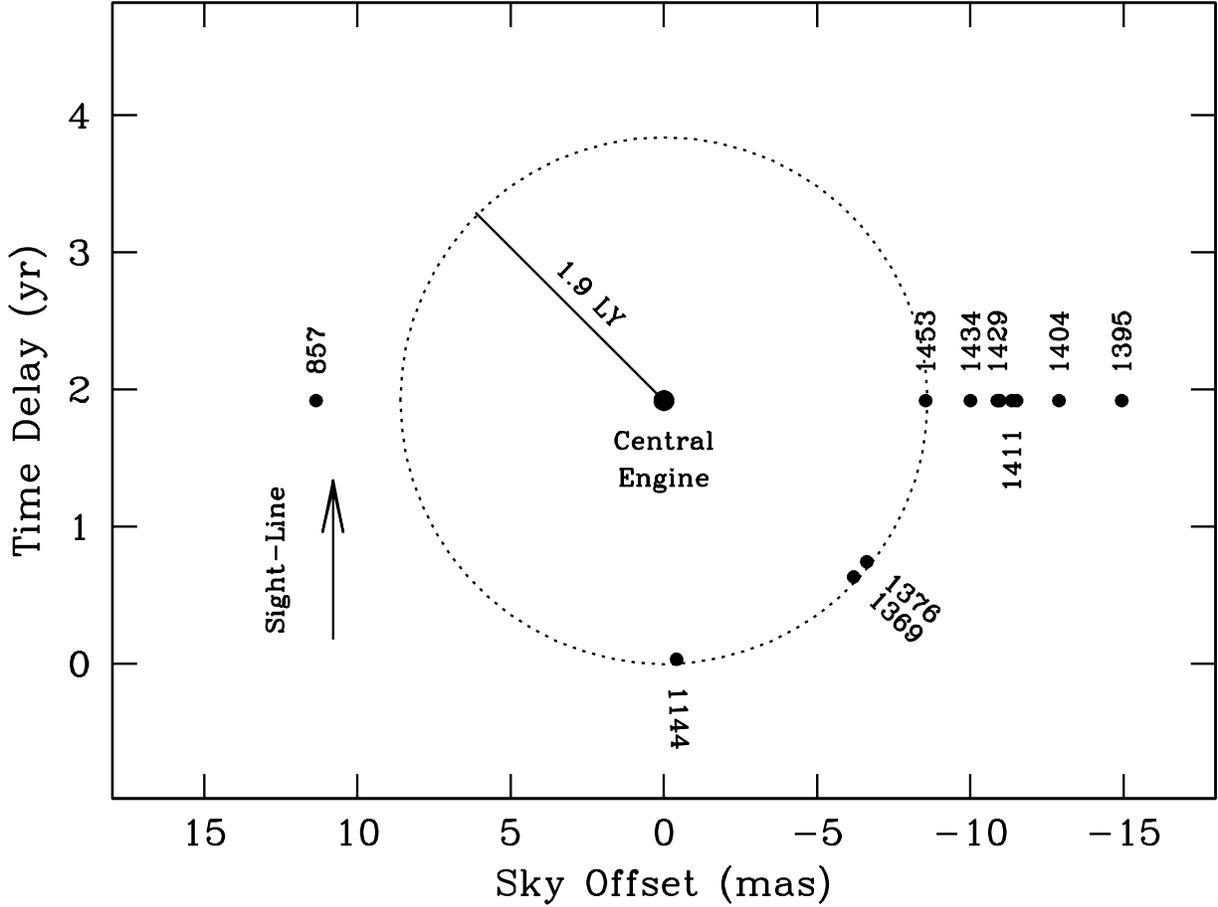}
\caption{The locations of the nuclear maser features assuming a
disk geometry. This schematic is a plan view looking down the
rotational pole of the disk.  The locations of the spots were
estimated based on the VLBA observations of Greenhill \& Gwinn (1997)
and a fit to their position-velocity curve. Each spot marks the
predicted location corresponding to a peak in the integrated maser
spectrum (see Table~\protect\ref{gausstab}). The spots are labeled by
their projected velocities in units \kms.  The geometry is derived by
setting the annular radius to the observed peak of the
position-velocity curve, or roughly 8.6~mas, and assuming that maser
spots on the falling portion of the position-velocity curve are
located along the disk midline. The reasoning is that the midline, or
line-of-nodes, offers the longest coherent light paths for maser
emission. It is also assumed that free-free absorption attenuates
maser emission from the far side of the disk. Note that bright
features of the integrated \h2o\ maser spectrum may include
contributions both from the inner, annular ring of masers and from
masers along the line of nodes; there appear, however to be no such
cases among the brighter maser features. The time-delay axis is
normalized assuming a distance of 14.4~Mpc to NGC~1068, with the
zero-point set at the location near-systemic masers. For example,
emission from the inner midline would lag behind emission from the
near-systemic masers by 1.9 years. }
\label{toyannulus}
\end{figure}

\begin{figure}\singlespace
\plotone{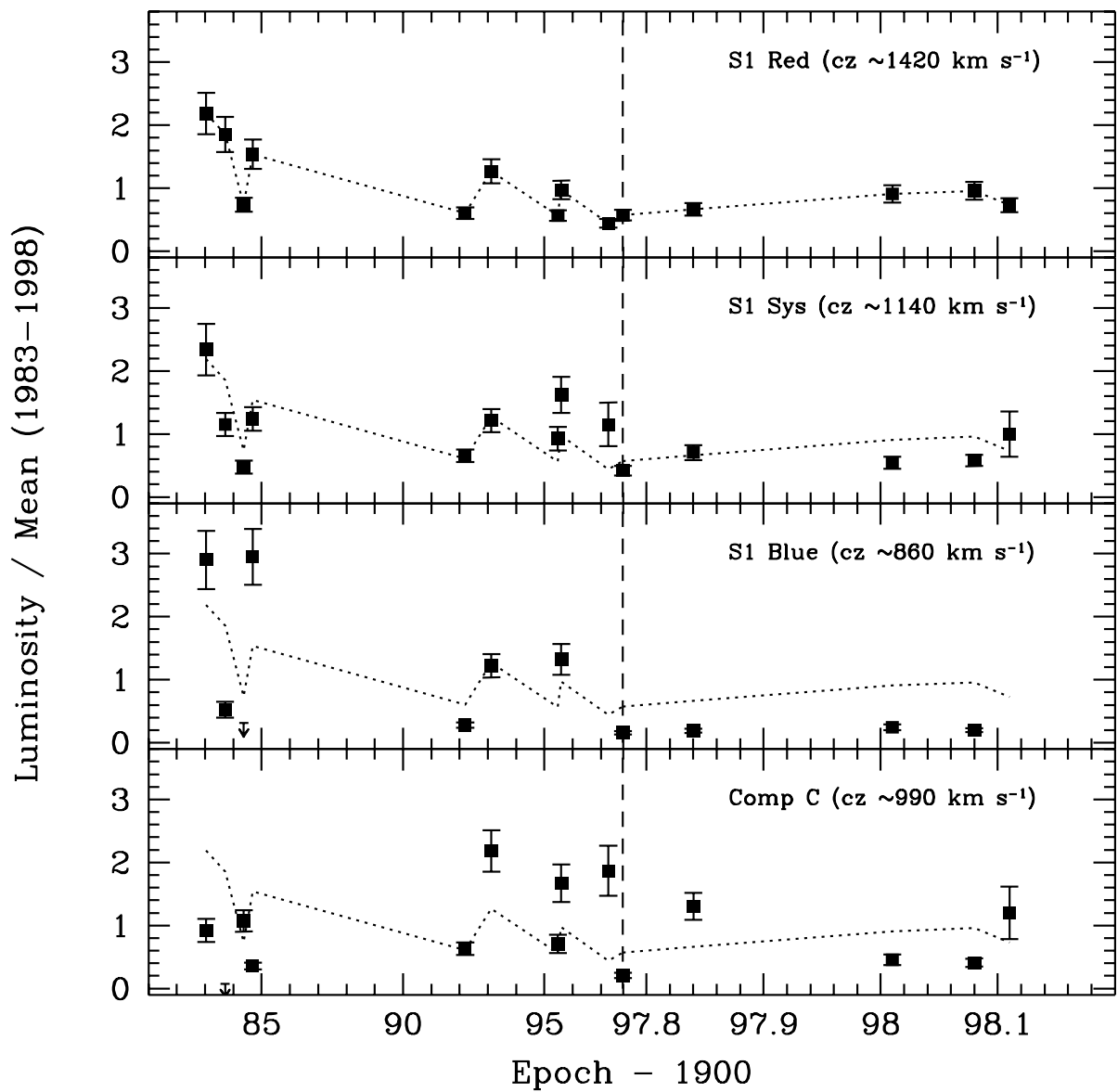}
\caption{Variations of the maser luminosities as a function of
observing epoch. For comparison with the bright red masers, all of the
maser luminosities are normalized to their mean over the monitoring
period. In each plot the dotted line traces the variations of the red
masers. ``Comp C'' refers to the jet masers at radio continuum
component C. To accommodate the more frequent sampling after 1996, we have
changed the scaling of the time axis at 1997.77 (the epoch where we
began more frequent monitoring), marked by the
vertical dashed line. }
\label{lumtime}
\end{figure}

\begin{figure}\singlespace
\plotone{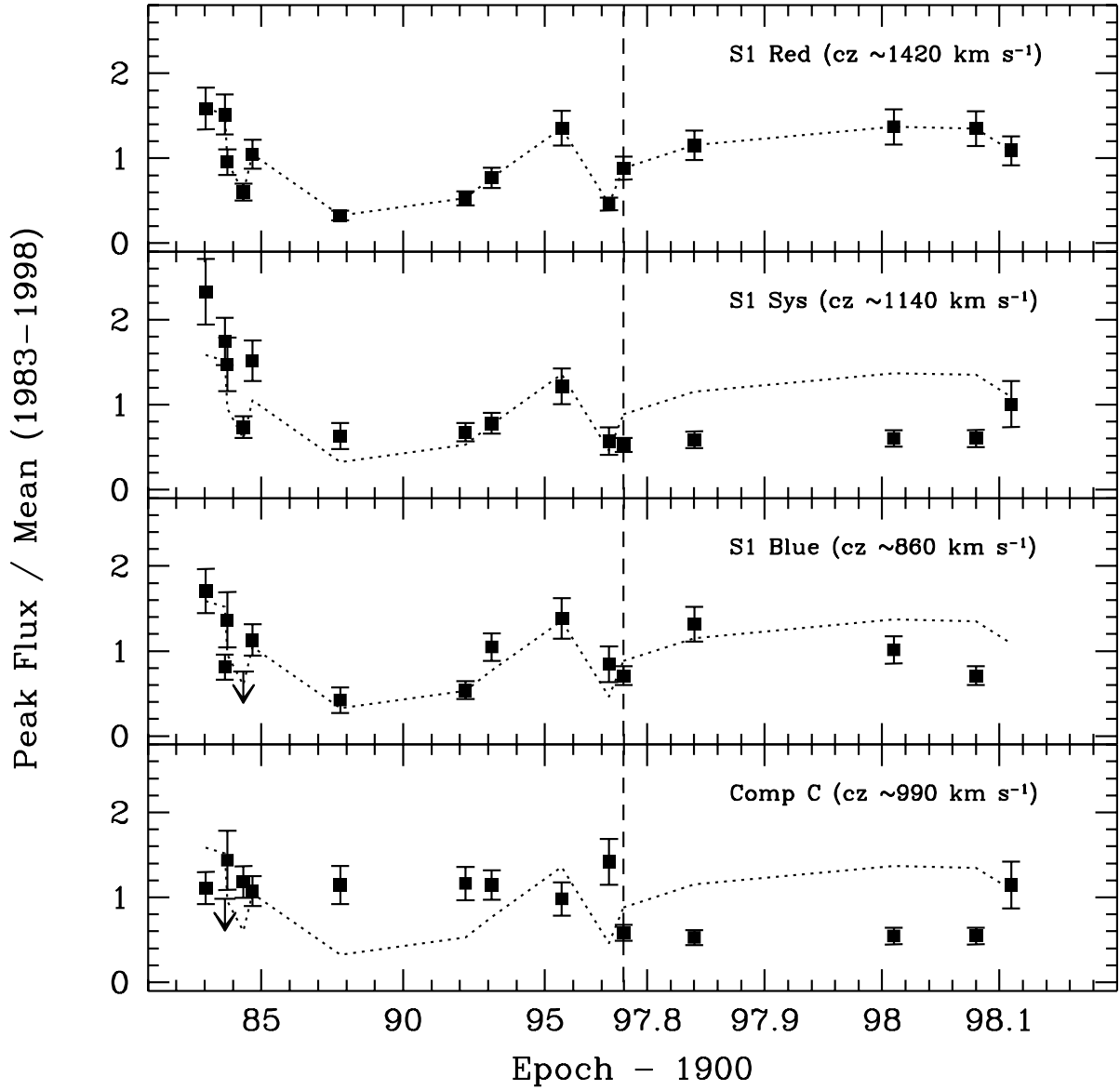}
\caption{Variations of the peak maser brightnesses as a function of
observing epoch. For comparison with the bright red masers, all of the
maser brightnesses are normalized to their mean over the monitoring
period. Otherwise the plotting convention is the same as for
Figure~\protect{\ref{lumtime}}.  } 
\label{fluxtime}
\end{figure}

\setcounter{figure}{11}
\begin{figure}\singlespace
\plotone{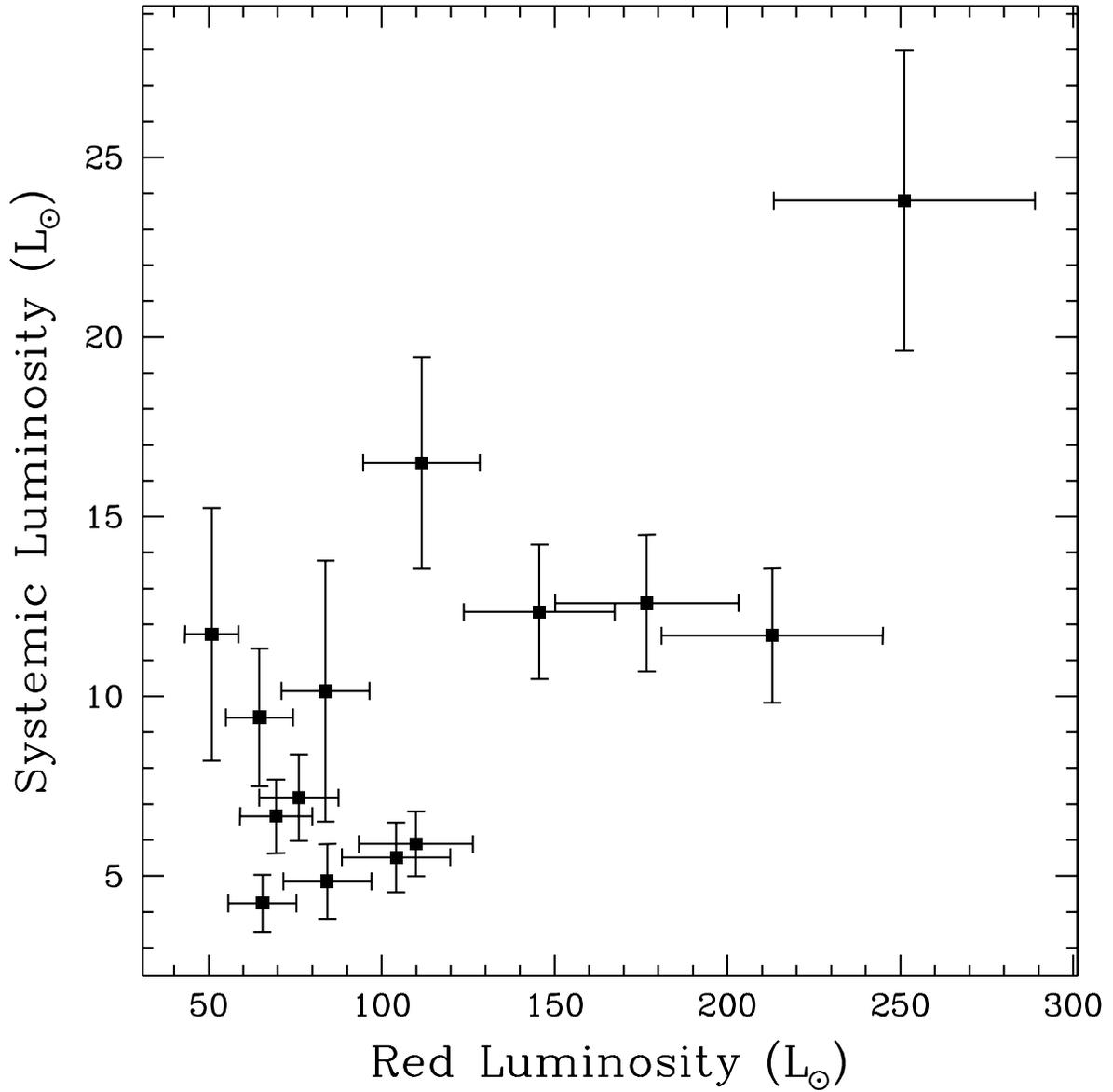}
\caption{(a) A plot of the luminosities of the near-systemic nuclear
masers against the luminosity of the red nuclear masers. The
reverberation model predicts little or no correlation between the red
and near-systemic masers. The errorbars include statistical
uncertainities and a relative scaling (flux calibration) error of 15\%. }
\label{lumlum}
\end{figure}

\setcounter{figure}{11}
\begin{figure}\singlespace
\plotone{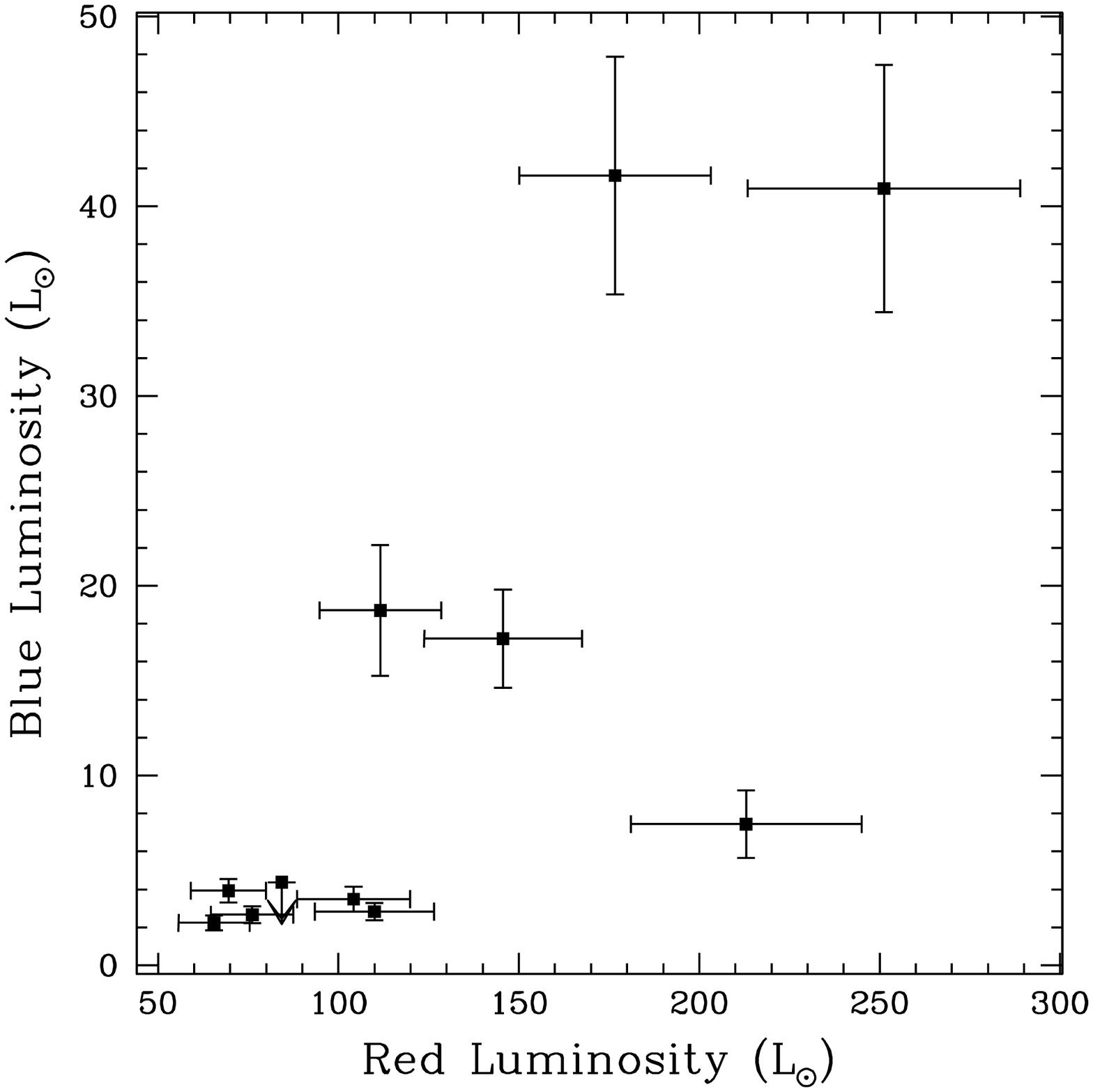}
\caption{(b) A plot of the luminosities of the blue nuclear
masers against the luminosity of the red nuclear masers. The
reverberation model predicts correlation of these (integrated)
luminosities if the pump variability timescale is longer than the signal travel
time through the maser disk.}
\label{lumlumb}
\end{figure}

\setcounter{figure}{11}
\begin{figure}\singlespace
\plotone{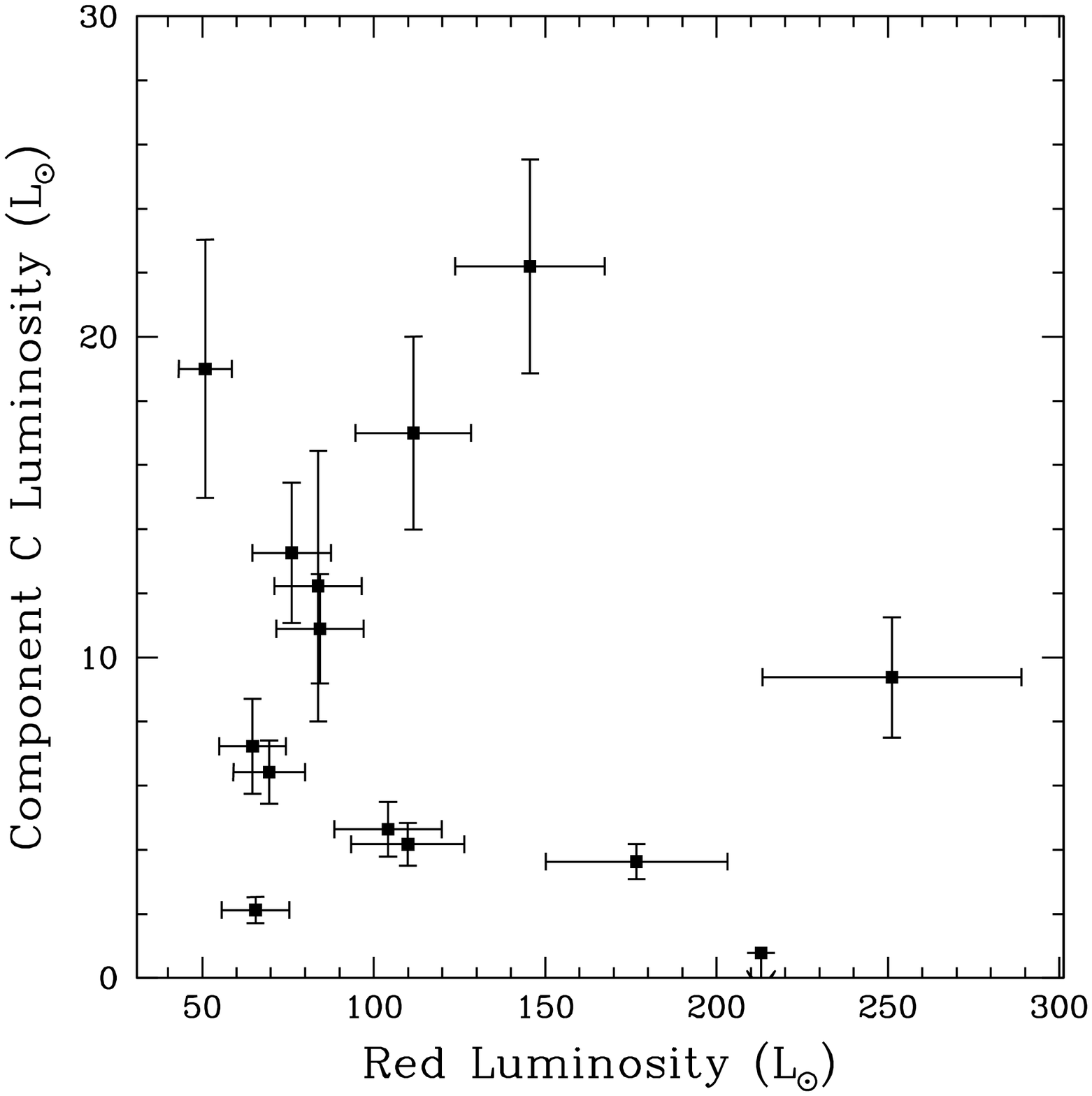}
\caption{(c) A plot of the luminosities of the jet (Component C)
masers against the luminosity of the red nuclear masers. Because the
separation between these maser sources is $\sim 30$~pc, the
reverberation model predicts no correlation between the red nuclear masers
and the jet masers. }
\label{lumlumc}
\end{figure}

\clearpage

\setcounter{figure}{12}
\begin{figure}\singlespace
\plotone{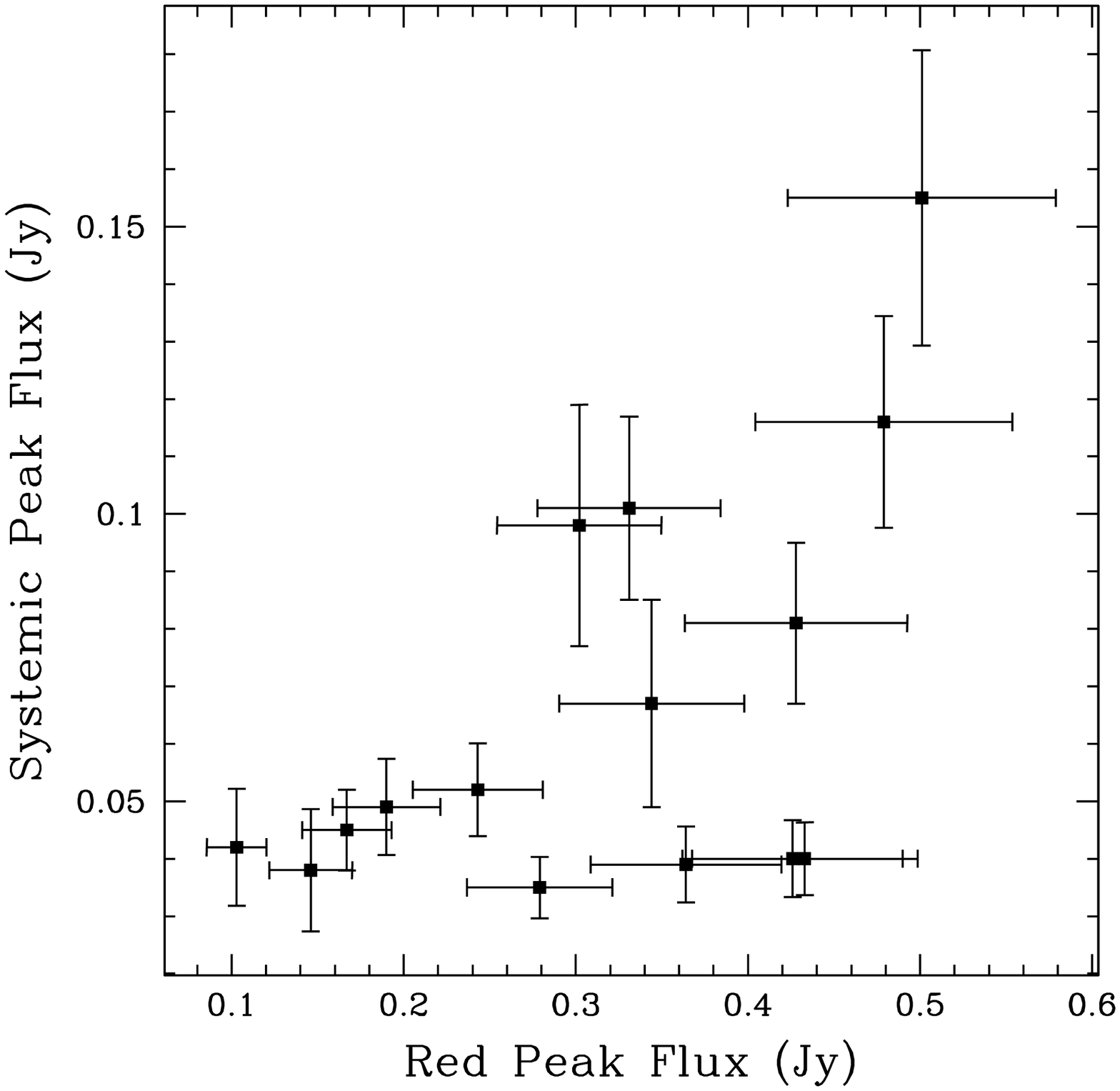}
\caption{(a) A plot of the peak brightnesses of the systemic nuclear
masers against the peak brightnesses of the red nuclear masers. The
reverberation model predicts little or no correlation between the
systemic and red masers.  The errorbars include statistical
uncertainities and a relative scaling (flux calibration) error of 15\%.}
\label{fluxflux}
\end{figure}

\setcounter{figure}{12}
\begin{figure}\singlespace
\plotone{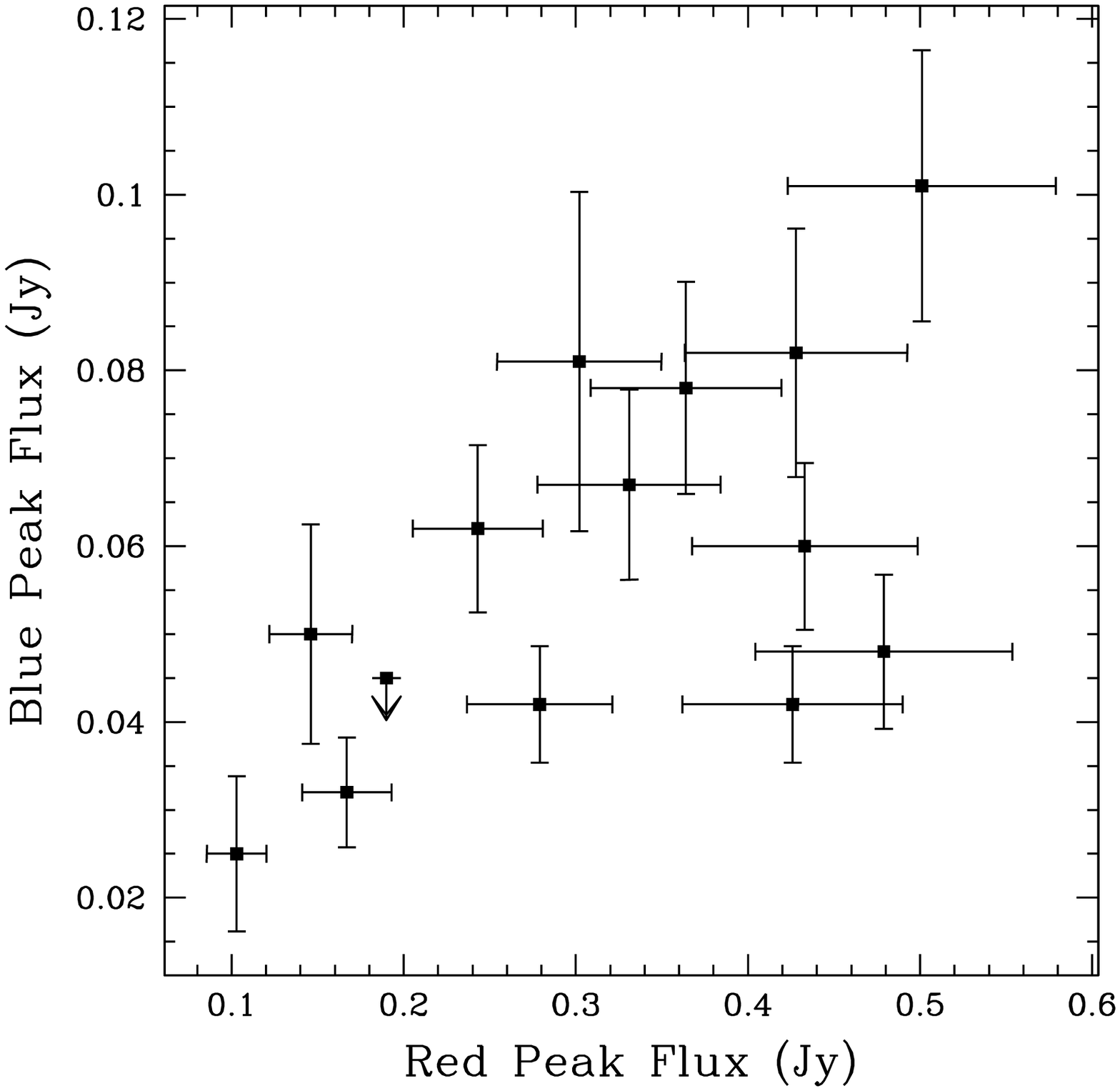}
\caption{(b) A plot of the peak brightnesses of the blue nuclear
masers against the peak brightnesses of the red nuclear masers. The
reverberation model predicts a correlation between the
blue and red masers.}
\label{fluxfluxb}
\end{figure}

\setcounter{figure}{12}
\begin{figure}\singlespace
\plotone{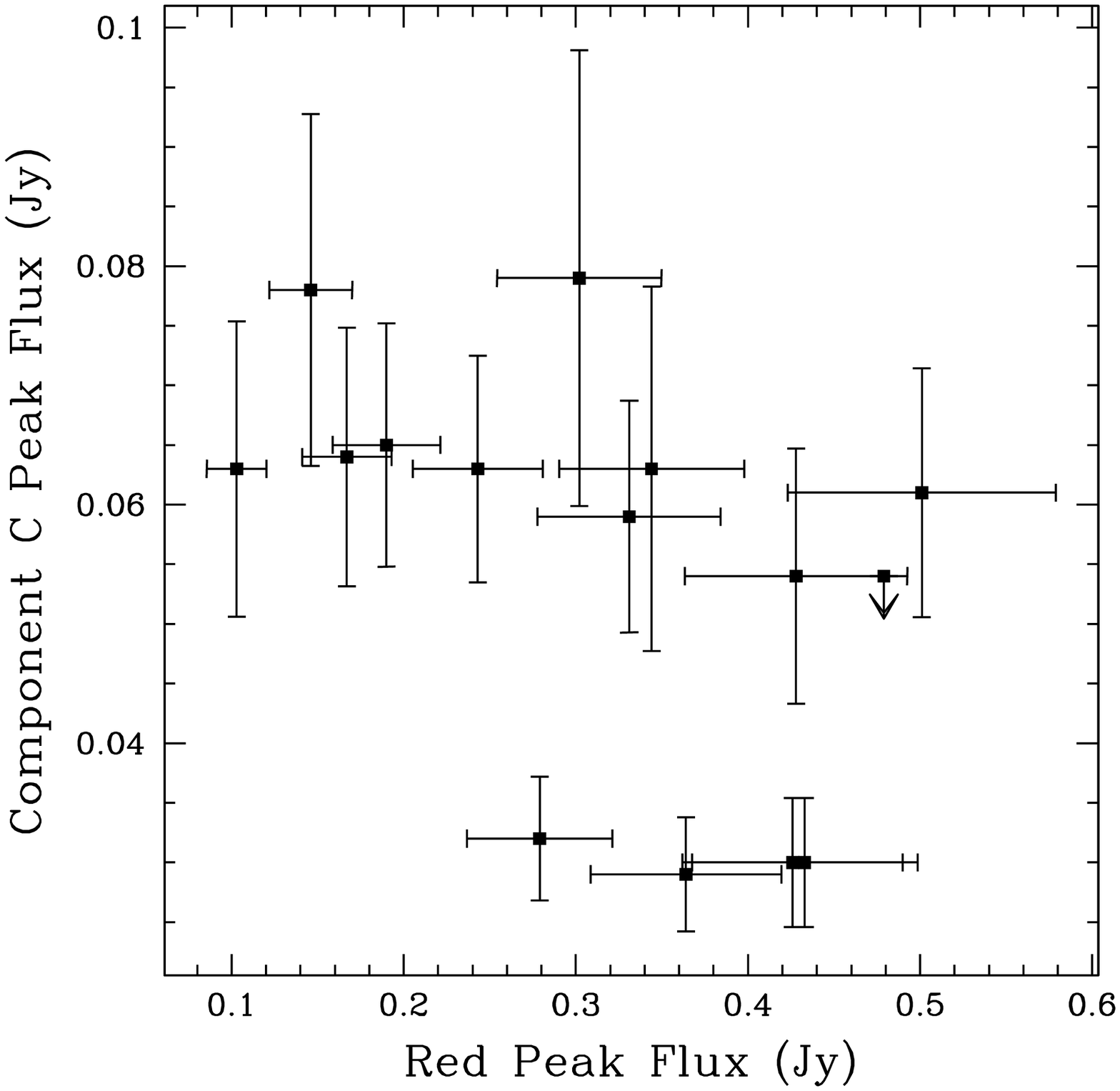}
\caption{(c) A plot of the peak brightnesses of the jet (``Component
C'') masers against the peak brightnesses of the red nuclear masers. The
reverberation model predicts no correlation between the
systemic and red masers.}
\label{fluxfluxc}
\end{figure}

\begin{figure}\singlespace
\plotone{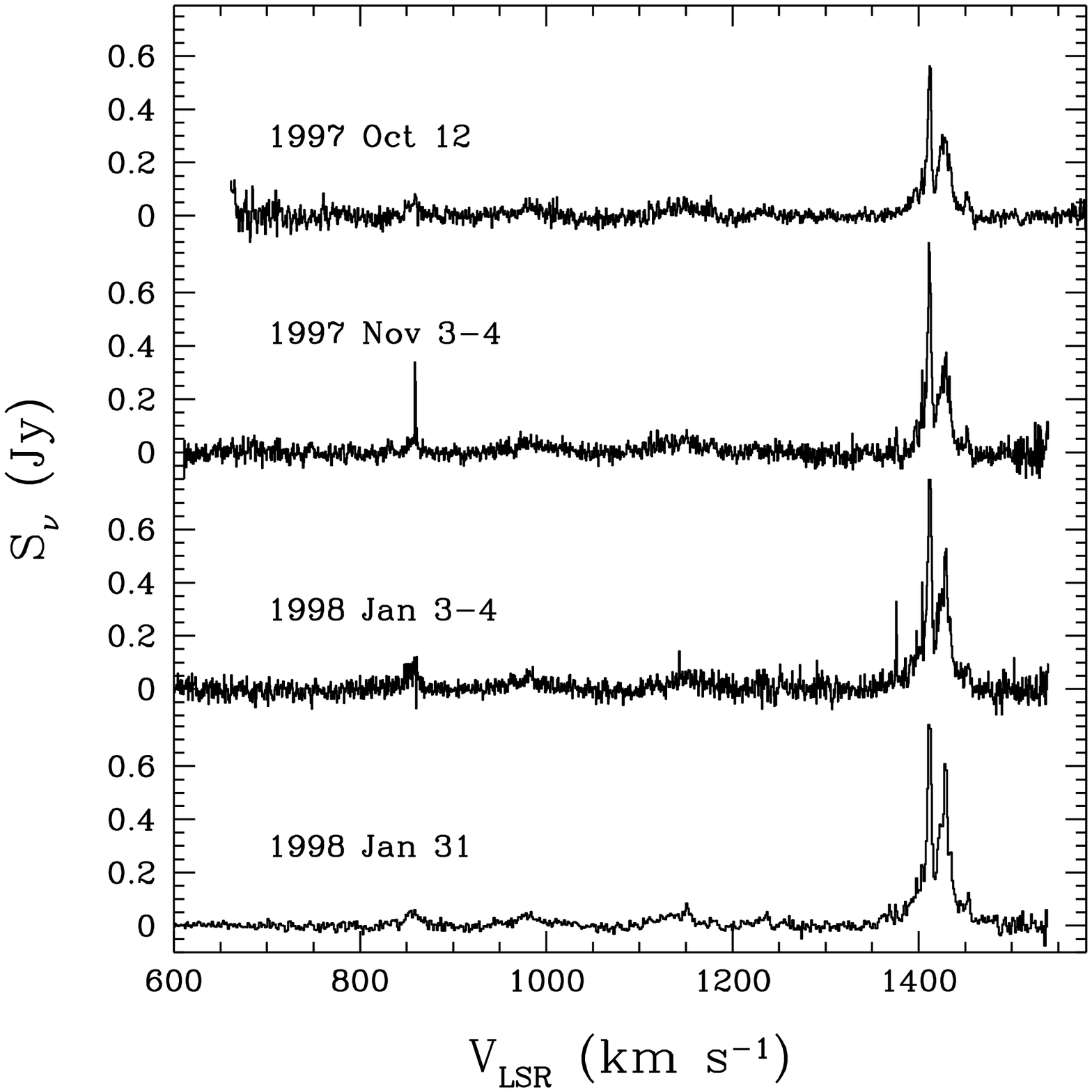}
\caption{History of the Nov. 1997 \h2o\ maser flare of
NGC~1068. There appears a simultaneous flare at 859~\kms\ and
1411~\kms\ in the Nov. 1997 data, and there is another flare at
1376~\kms\ in the 3 Jan 1998 data.}
\label{maserflare}
\end{figure}

\begin{figure}\singlespace
\plotone{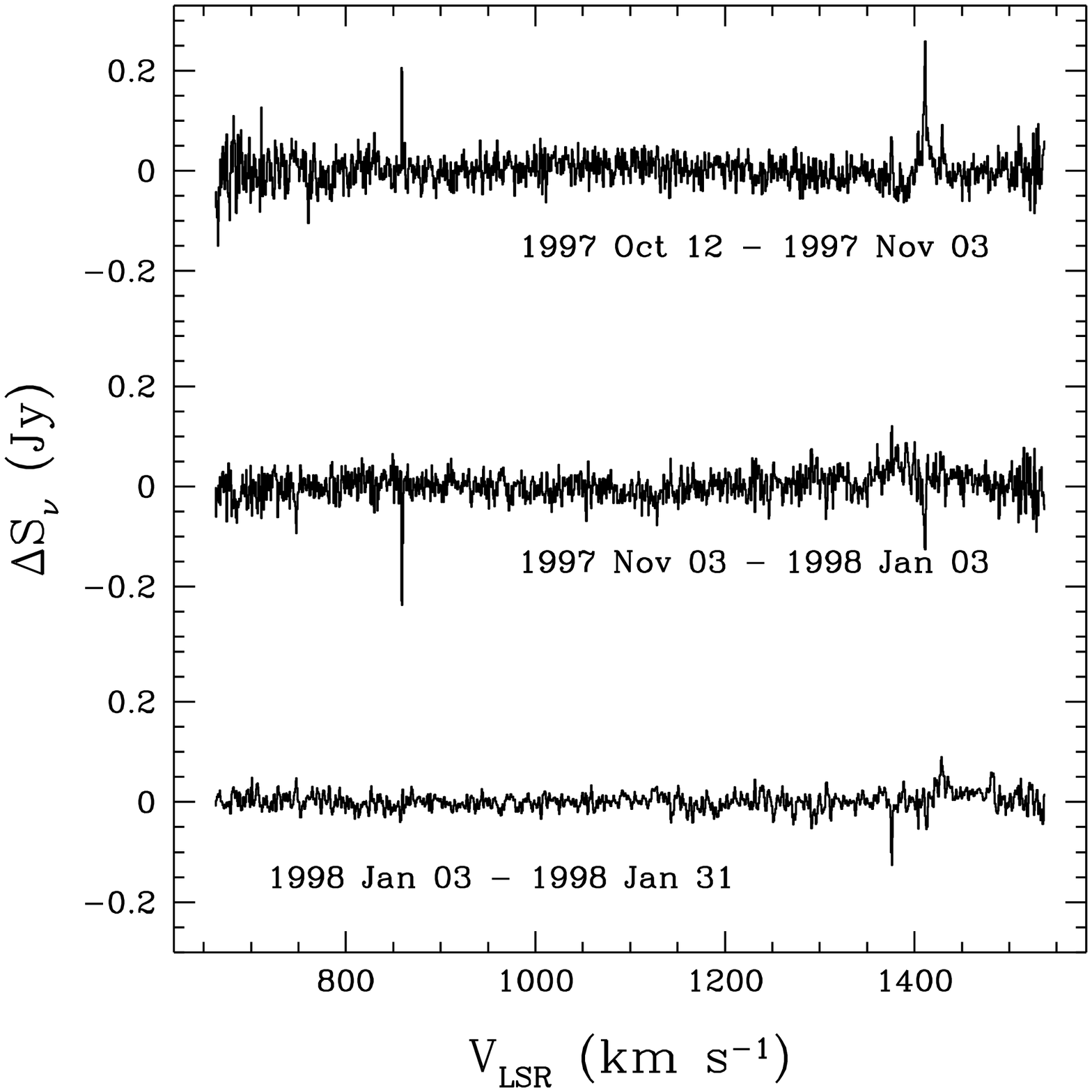}
\caption{History of the Nov. 1997 \h2o\ maser flare of
NGC~1068. Plotted are the difference spectra between successive
observations. In computing the difference spectra, the individual
spectra were scaled to minimize the rms of the off-flare spectra. The
purpose was to try to account for absolute calibration errors between
successive observations. The scaling factors are $< 20\%$, within the
range of expected flux calibration uncertainty. }
\label{maserflare2}
\end{figure}

\begin{figure}\singlespace
\plotone{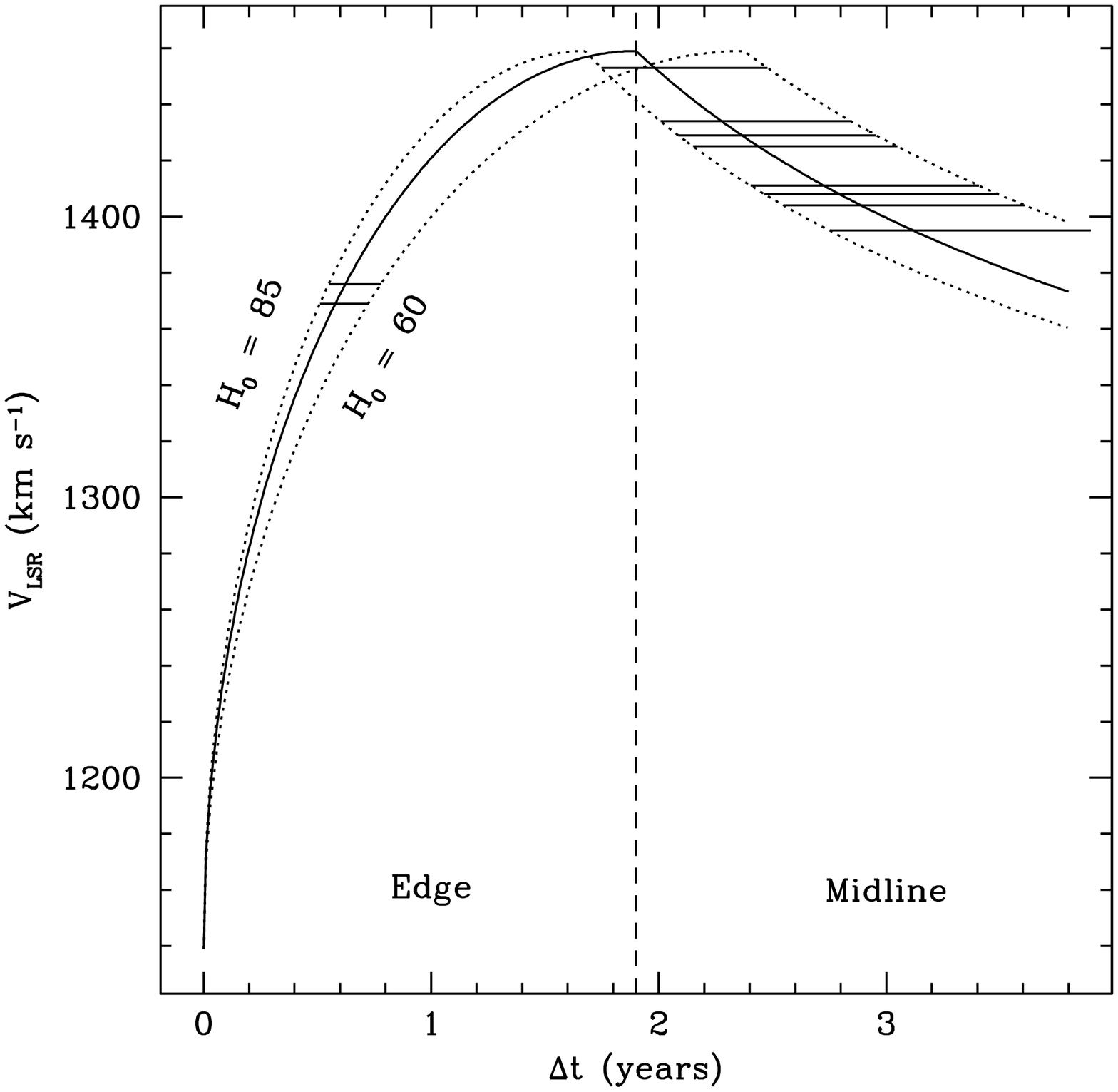}
\caption{Predicted velocity for nuclear \h2o\ maser flares as a
function of time-lag. The assumed geometry is a rotating annulus.  The
solid curve traces the prediction for NGC~1068, assuming a distance
$D=14.4 $~Mpc, which gives the size of the maser annulus $R=0.6$~pc,
and rotation velocity $v_{rot}$ = 320~\kms. Zero time-lag is defined
at the systemic velocity $= 1139$~\kms. The solid, horizontal lines
mark the velocities of the brightest maser features on the single dish
spectra. We estimated the time-lags for these velocities based on the
plots of Greenhill \& Gwinn (1997). The vertical, dashed line
separates the response of the masers along the inner edge of the
annulus and the masers along the midline (line-of-nodes). The dotted
lines trace the velocity-lag curves for distances of 12.7 Mpc ($H_0 =
85$~\kms\ Mpc\mone) and 18 Mpc ($H_0 = 60$~\kms\ Mpc\mone).  }
\label{drifts}
\end{figure}

\begin{figure}\singlespace
\plotone{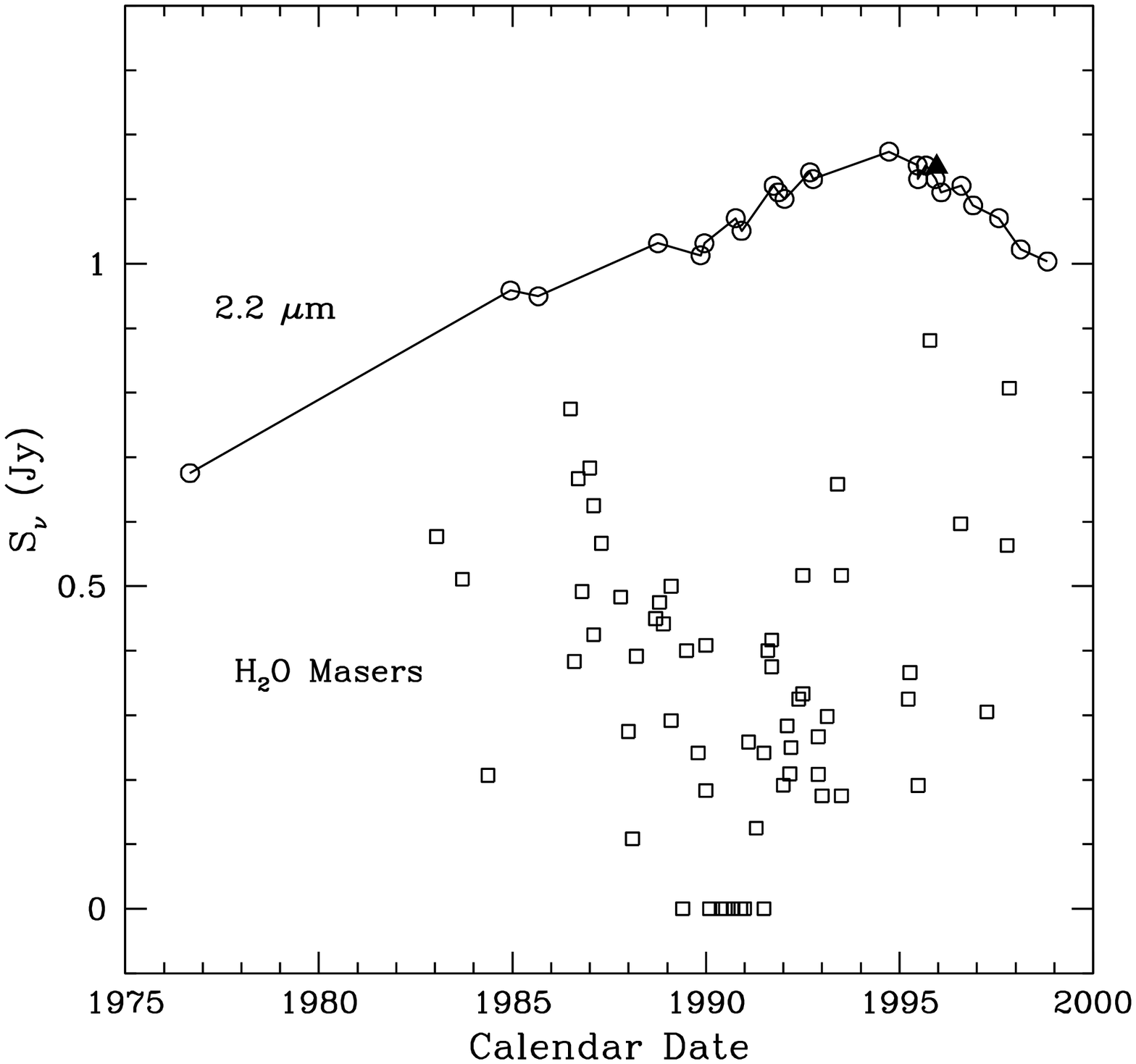}
\caption{Comparison of the time variability of the near-infrared
K-band continuum (upper curve defined by open circles), from Glass
(1997), and the 22~GHz \h2o\ maser emission (defined by
open squares). The filled triangle marks the measurement of Weinberger
et al. (1999). The maser emission curve plots the peak brightness of
the 1411~\kms\ feature; most of the densely sampled data around 1990
are taken from Baan \& Haschick 1996. We suggest that the peak of the
2.2\micron\ curve might be the time-lagged response of hot dust
emission to the same X-ray flare that may have extinguished the \h2o\
masers in 1990.}
\label{irplot}
\end{figure}

\begin{figure}\singlespace
\plotone{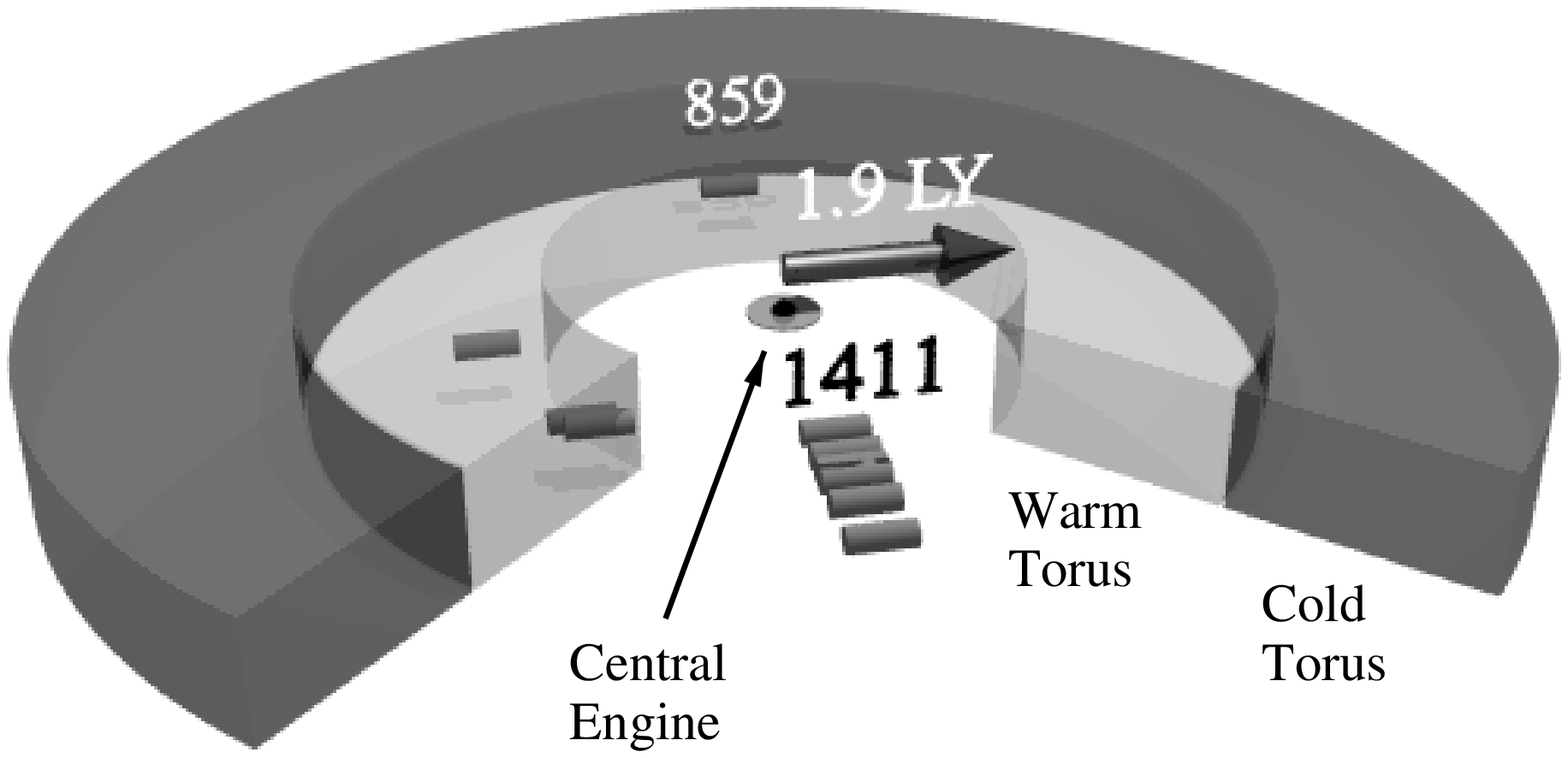}
\caption{A schematic illustrating the geometry of the \h2o\ maser
torus based on the maser-IR reverberation model described in
\S\protect\ref{ircraziness}. This figure illustrates a cutaway view
of the torus, corresponding to the plan view plotted in
Figure~\ref{toyannulus}. The ``warm torus'' region describes the
location of both the \h2o\ masers, rendered as dark cylinders, and hot
dust on the inner surface of the obscuring disk. The locations of the
859~\kms\ and 1411~\kms\ masers are labeled above the corresponding
dark cylinder. The ``cold torus'' describes a region of cooler dust
where the \h2o\ abundance should be reduced (e.g., Maloney \& Neufeld
1995).  The dark arrow indicates the radius to the inner surface of
the maser annulus traced along our sight-line.  The time-lag between
the loss of red maser signal and the peak of the near-infrared light
curve may result partly because hot dust in the near-side of the warm
torus is obscured by the cold torus. We see a partially eclipsed view
of the far side of the warm torus. Distances are normalized assuming
circular symmetry and that the distance to NGC~1068 is 14.4~Mpc. }
\label{irschematic}
\end{figure}

\begin{figure}\singlespace
\plotone{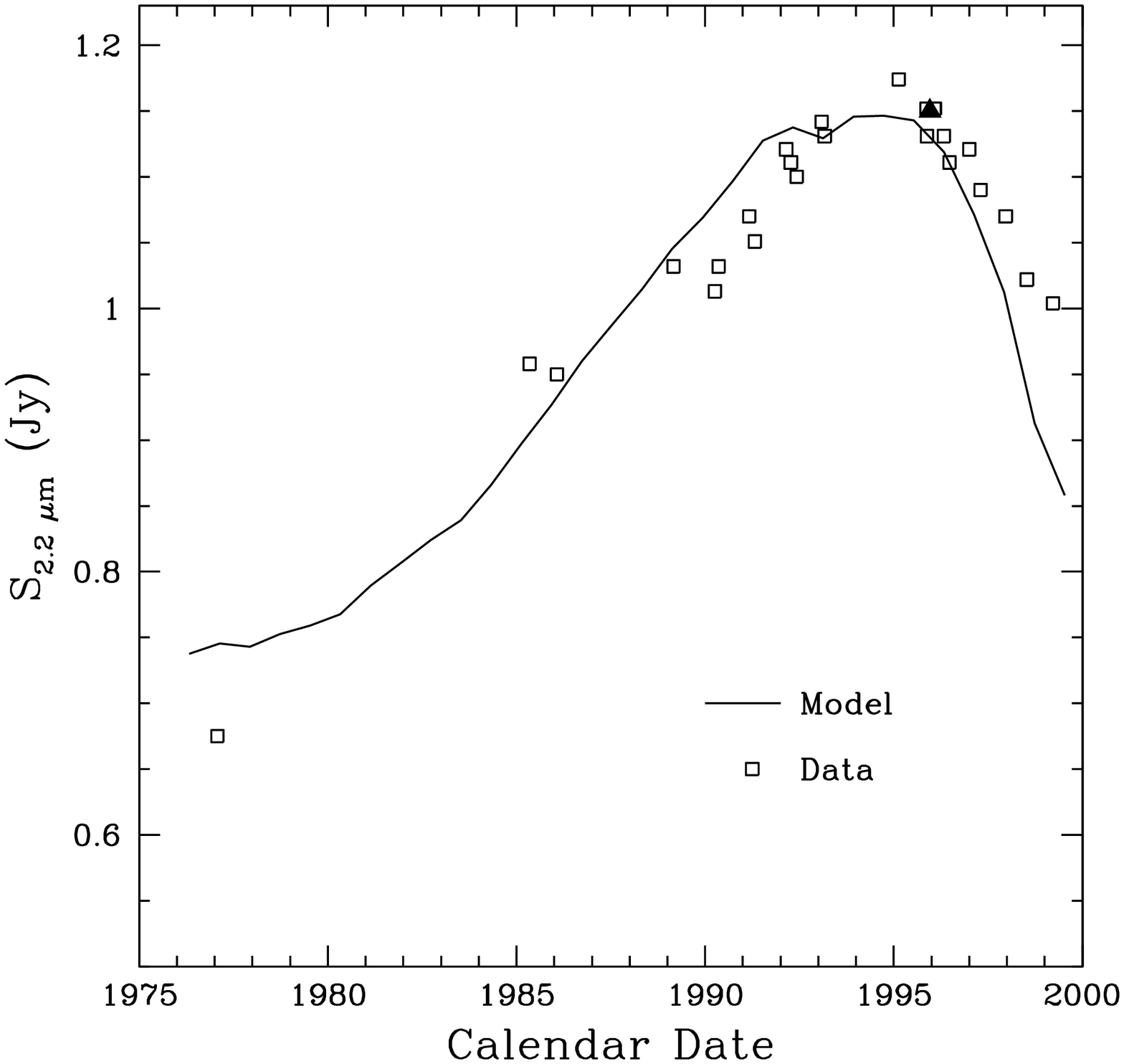}
\caption{The response of hot dust emission to an X-ray flare of the
central engine. The model is that  
described in \S\protect\ref{irmodel}, and illustrated in
Figure~\protect\ref{irschematic}. The solid line traces the best-fit
model, the filled squares mark the
monitoring observations of Glass (1997), and the filled triangle marks
the measurement of Weinberger et al. (1999). The parameters of the model
are listed in Table~\protect\ref{t_irmodel}. 
}\label{irresponse}
\end{figure}


%% file: tables.tex
\begin{deluxetable}{rrrrr@{--}l}
\singlespace
\tablewidth{0pc}
\tablecaption{Effelsberg 100m Observations Log}
\tablehead{
\mccc{Date} & \mc{$\Delta v$} & \mcc{Velocity Range} \\
 & & & \mc{(\protect{\kms})} & \mcc{(LSR, \protect{\kms})} \\
\mccc{(1)} & \mc{(2)} & \mcc{(3)} 
}
\startdata
1984 & Sep & 5-6 &   1.33 &  1000 &  1600 \\
1985 & Jul &  13 &   0.66 &  1280 &  1520 \\
1985 & Aug &   4 &   0.66 &  1280 &  1520 \\
1985 & Sep &  14 &   0.66 &  1000 &  1600 \\
1995 & Jan &  15 &   1.33 &   860 &  1588 \\
1995 & May &  10 &   0.66 &   860 &  1759 \\
1995 & Jun &  26 &   0.33 &  1230 &  1590 \\
1995 & Sep &  16 &   0.33 &  1230 &  1570 \\
1996 & Jul &  30 &   0.33 &  1315 &  1484 \\
1997 & Apr &   6 &   0.66 &   860 &  1540 \\
1997 & Oct &  12 &   0.66 &   660 &  1620 \\
1997 & Nov & 3--4&   0.66 &   610 &  1540 \\
1998 & Jan &   3 &   0.66 &   460 &  1540 \\
1998 & Jan &  31 &   1.33 &   560 &  1540 \\
1998 & Feb &  10 &   1.33 &   860 &  1540 \\
1998 & May &   8 &   0.33 &  1314 &  1484 
\tablecomments{Columns are (1) observing date, (2) channel width, and
(3) velocity range covered by the observations. Note that we usually
employed several overlapping tunings to cover the reported velocity
ranges, and, as a result, the noise levels are not uniform across
these velocity ranges. }\label{obstable}
\enddata
\end{deluxetable}

\begin{deluxetable}{rrll}
\singlespace
\tablewidth{0pc}
\tablecaption{Bibliography of Published \h2o\ Maser Spectra of NGC~1068}
\tablehead{
\mcc{Obs. Date} & \mc{Telescope} & \mc{Reference} \\
\mcc{(1)} & \mc{(2)} & \mc{(3)} 
}
\startdata
1983 & Jan      & OVRO 40m & Claussen \& Lo 1986\\
1983 & Sep      & OVRO 40m & Claussen \& Lo 1986\\
1984 & May      & OVRO 40m & Claussen \& Lo 1986\\
1984 & Sep & Effelsberg 100m & Henkel et al. 1984 \\
1992 & Jan--Jun & Nobeyama 45m & Nakai et al. 1995 \\
1993 & Feb & Effelsberg 100m & Greenhill et al. 1996 
\tablecomments{Columns are (1) observing date, (2) telescope used for
observations, and (3) reference for the published spectrum.}\label{pubspectra}
\enddata
\end{deluxetable}

\nocite{HGWBDT84}
\nocite{GGAB96}

\begin{deluxetable}{lrr}
\singlespace
\tablewidth{0pc}
\tablecaption{\h2o\ Maser Velocities}
\tablehead{
 Radio &  \mcc{$v_{LSR}$} \\
 ID    &  \mcc{(\kms)}    
}
\startdata
S1 &  856.6 & $\pm$0.3 \\
C  &  984.1 & 1.2 \\
S1 & 1143.9 & 1.1 \\
S1 & 1368.5 & 0.2 \\
S1 & 1376.1 & 0.1 \\
S1 & 1394.8 & 0.7 \\
S1 & 1403.8 & 0.1 \\
S1 & 1407.5 & 0.3 \\
S1 & 1411.6 & 0.7 \\
S1 & 1424.6 & 0.3 \\
S1 & 1429.3 & 0.1 \\
S1 & 1434.0 & 0.3 \\
S1 & 1452.9 & 0.2 
\tablecomments{The LSR velocities of the \h2o\ masers of
NGC~1068. The first columns lists the nearest radio continuum
component to the maser feature as identified by VLA imaging. Only
those features that are easily distinguishable at 0.66~\kms\ channel
spacing are listed. The velocities were measured from local Gaussian
fits to the line profile in the integrated maser spectrum derived from
1995--1998 Effelsberg 100m observations.  }\label{gausstab}
\enddata
\end{deluxetable}

\begin{deluxetable}{llrrrrrrrr}
\singlespace
\tablewidth{0pc}
\tablecaption{Maser Variability Correlation Statistics }
\tablehead{
\mc{Experiment}  & \mcc{Test on Peaks} & \mc{Prob}  & \mc{Prob$_{MC}$} & \mc{$n$} 
                 & \mc{Test on Lums.} & \mc{Prob}  & \mc{Prob$_{MC}$}  & \mc{$n$} \\
\mc{(1)}         & \mcc{(2)}           & \mc{(3)}   & \mc{(4)}          & \mc{(5)} 
                 & \mc{(6)}           & \mc{(7)}   & \mc{(8)}          & \mc{(9)} 
}
\startdata
Blue on Red 
            & Pearson $r$       &   0.728 & 0.016 & 0.040 & 10 & 0.729 & 0.017 & 0.042 & 10 \\
            & Spearman $r  $    &   0.806 & 0.005 & 0.006 & 10 & 0.806 & 0.005 & 0.006 & 10 \\
            & Cox $\chi^2  $    &  10.264 & 0.001 & 0.003 & 11 &10.176 & 0.001 & 0.010 & 11 \\
            & Kendall $\tau  $  &   2.569 & 0.010 & 0.008 & 11 & 2.569 & 0.010 & 0.007 & 11 \\
\tableline
Systemic 
            & Pearson $r$       &   0.603 & 0.013 & 0.012 & 16 & 0.704 & 0.005 & 0.010 & 14 \\
on Red      & Spearman $r  $    &   0.515 & 0.041 & 0.040 & 16 & 0.526 & 0.053 & 0.056 & 14 \\
            & Cox $\chi^2  $    &   3.589 & 0.058 & 0.068 & 16 & 7.772 & 0.005 & 0.014 & 14 \\
            & Kendall $\tau$    &   2.208 & 0.027 & 0.023 & 16 & 1.806 & 0.071 & 0.066 & 14 \\
\tableline
Jet on Red 
            & Pearson $r$       &  $-$0.150 & 0.593 & 0.602 & 15 & $-$0.031 & 0.919 & 0.944 & 13 \\
            & Spearman $r  $    &  $-$0.205 & 0.464 & 0.462 & 15 & $-$0.016 & 0.957 & 0.964 & 13 \\
            & Cox  $\chi^2  $   &   2.212 & 0.137 & 0.157 & 16 &  0.135 & 0.712 & 0.690 & 14 \\
            & Kendall  $\tau  $ &  $-$0.769 & 0.441 & 0.453 & 16 & $-$0.164 & 0.869 & 0.828 & 14 
\tablecomments{Column 1 lists the maser groups under consideration
(e.g., ``Blue on Red'' means correlation statistics between the
nuclear blue and red masers). The next four columns (2--5) list the
correlation statistics comparing the peak flux of the maser spectrum
for those groups of masers, the formal probability for no correlation,
the same probability estimated using Monte Carlo statistics
(Prob$_{MC}$), and the number of data points used in the test. A low
probability means there is evidence for significant
correlation. Columns 6--8 list the corresponding statistics comparing
the integrated luminosities of the maser groups. Note that the data do
not always completely cover the maser emission spectrum, and, as a
result, there are usually fewer luminosity measurements than peak
measurements.  }\label{corrtab}
\enddata
\end{deluxetable}

\begin{deluxetable}{lrr}
\singlespace
\tablewidth{0pc}
\tablecaption{Dust Reverberation Model Parameters}
\tablehead{
\mc{Parameter} & \mc{Value} & \mc{Error} 
}
\startdata
Minimum Luminosity     & $1.1\times 10^{11}~L_{\sun}$ & $0.5\times 10^{11}~L_{\sun}$\\
Maximum Luminosity     & $5.2\times 10^{11}~L_{\sun}$ & $0.8\times 10^{11}~L_{\sun}$\\
Flare Duration         & 9.8~years & 2.4~years\\
Hydrogen Density       & $4.0\times 10^{6}$~cm\mthree\ & $1.5 \times 10^6$~cm\mthree \\
Disk Inclination       & 87\arcdeg & 4\arcdeg \\
Inner Edge Temperature & 1300~K & (Fixed) 
\tablecomments{The reported errors were estimated based on the
variance of the fitted parameters with changes of the model
systematics. To this end we adjusted (1) initial guesses of the model
parameters; (2) the shape of the flare profile (Gaussian, sawtooth,
sawtooth plus slope, and tophat); (3) the fraction of starlight in the
12\arcsec\ aperture; and (4) the assumed disk scale height. The
reported flare duration is the FWHM of the model flare. The hydrogen
density was computed assuming a standard gas-to-dust ratio and
extinction curve \citep{DL84}.}\label{t_irmodel}
\enddata
\end{deluxetable}

%% file: ms.bbl
\begin{thebibliography}{}
\bibitem[{Antonucci \& Miller}(1985)]{AM85} Antonucci, R. R. J. \&
Miller, J. S. 1985, \apj, 297, 621
\bibitem[{Baan \& Haschick}(1996)]{BH96} Baan, W. A. \& Haschick,
A. 1996, \apj, 473, 269
\bibitem[{Baars et al.}(1977)]{BGPW77} Baars, J. W. M., Genzel, R.,
Pauliny-Toth, I. I. K., \& Witzel, A. 1977, \aap, 619, 99
\bibitem[{Barvainis}(1992)]{Barvainis92} Barvainis, R. 1992, \apj,
400, 502
\bibitem[{Begelman \& Bland-Hawthorn}(1997)]{BB97} Begelman,
M. C. \& Bland-Hawthorn, J. 1997, \nat, 385, 22
\bibitem[{Bode}(1988)]{Bode88} Bode, M. F., in Dust in the Universe,
eds. M. E. Baily \& D. A. Williams (Cambridge: Cambridge University
Press), 73  
\bibitem[{Bragg et al.}(2000)]{Bragg00} Bragg, A. E., Greenhill,
L. J., Moran, J. M., \& Henkel, C. 2000, \apj, 535, 73
\bibitem[{Brinks et al.}(1997)]{BSTT97} Brinks, E., Skillman, E. D.,
Terlevich, R. J., \& Terlevich, E. 1997, \apss, 248, 23
\bibitem[{Capetti et al.}(1995)]{Capetti95b} Capetti, A., Macchetto,
F., Axon, D. J., Sparks, W. B., \& Boksenberg, A. 1995, \apjl, 452,
L87
\bibitem[{Capetti, Macchetto, \& Lattanzi}(1997)]{CML97} Capetti,
A., Macchetto, F. D., \& Lattanzi, M. G. 1997, \apj, 476, L67
\bibitem[{Claussen, Heiligman, \& Lo}(1984)]{CHL84} Claussen, M. J.,
Heiligman, G. M., \& Lo, K. -Y. 1984, \nat, 310, 298
\bibitem[{Claussen \& Lo}(1986)]{CL86} Claussen, M. J. \& Lo,
K. -Y. 1986, \apj, 308, 592
\bibitem[{Claussen et al.}(1996)]{CWBWMT96} Claussen, M. J., Wilking,
B. A., Benson, P. J., Wootten, A., Myers, P. C., \& Terebey, S. 1996,
\apjs, 106, 111
\bibitem[{Clavel, Wamsteker, \& Glass}(1989)]{CWG89} Clavel, J.,
Wamsteker, W., \& Glass, I. S. 1989, \apj, 337, 236
\bibitem[{Dopita \& Sutherland}(1995)]{DS95b} Dopita, M. A. \&
Sutherland, R. S. 1995, \apj, 455, 468
\bibitem[{Draine \& Lee}(1984)]{DL84} Draine, B. T., \& Lee,
H. M. 1984, \apj, 285, 89
\bibitem[{Elitzur, Hollenbach, \& McKee}(1989)]{EHM89} Elitzur, M.,
Hollenbach, D. J., \& McKee, C. F. 1989, \apj, 346, 983
\bibitem[{Evans et al.}(1991)]{EFKAAC91} Evans, I. N., Ford, H. C.,
Kinney, A. L., Antonucci, R. R. J., Armus, L., \& Caganoff, S. 1991,
\apjl, 369, L27
\bibitem[{Fomalont}(1994)]{Fomalont94p219} Fomalont,
E. B. 1994, Synthesis Imaging in Radio Astronomy, ed. R. A. Perley,
F. R. Schwab, \& A. Bridle, (San Francisco: A. S. P.), 219
\bibitem[{Gallimore}(1996)]{mythesis} Gallimore,
J. F. 1996, Ph. D. Dissertation (College Park: University of Maryland)
\bibitem[{Gallimore, Baum, \& O'Dea}(1997)]{naturepaper} Gallimore,
J. F., Baum, S. A., \& O'Dea, C. P. 1997, \nat, 388, 852
\bibitem[{Gallimore et al.}(1996)]{maserpaper} Gallimore, J. F.,
Baum, S. A., O'Dea, C. P., Brinks, E., \& Pedlar, A. 1996, \apj, 462,
740
\bibitem[{Gallimore et al.}(1997)]{Kyoto} Gallimore, J. F., Baum,
S. A., O'Dea, C. P., \& Claussen, M. 1997, The AGN-Megamaser
Connection, IAU Joint Discussion 21, 26 (Kyoto: 23rd IAU General Assembly)
\bibitem[{Gallimore, Baum, \& O'Dea}(1996)]{n1068b} Gallimore,
J. F., Baum, S. A., \& O'Dea, C. P. 1996, \apj, 464, 198
\bibitem[{Gallimore et al.}(1996)]{n1068a} Gallimore, J. F., Baum,
S. A., O'Dea, C. P., \& Pedlar, A. 1996, \apj, 458, 136
\bibitem[{Glass}(1997)]{Glass97} Glass, I. S. 1997, \apss, 248, 191
\bibitem[{Goldreich \& Kwan}(1974)]{GK74} Goldreich, P. \& Kwan,
J. 1974, \apj, 191, 93
\bibitem[{Greenhill \& Gwinn}(1997)]{GG97} Greenhill, L. J. \&
Gwinn, C. R. 1997, \apss, 248, 261
\bibitem[{Greenhill et al.}(1996)]{GGAB96} Greenhill, L. J., Gwinn,
C. R., Antonucci, R., \& Barvainis, R. 1996, \apjl, 472, L21
\bibitem[{Henkel et al.}(1984)]{HGWBDT84} Henkel, C., Guesten, R.,
Wilson, T. L., Biermann, P., Downes, D., \& Thum, C. 1984, \aap, 141, L1
\bibitem[{Herrnstein et al.}(1999)]{HMGDINMHR99} Herrnstein, J. R.,
et al. 1999, \nat, 400, 539
\bibitem[{Isobe, Feigelson, \& Nelson}(1986)]{IFN86} Isobe, T.,
Feigelson, E. D., \& Nelson, P. I. 1986, \apj, 306, 490
\bibitem[{Jones et al.}(1994)]{JTHM94} Jones, A. P., Tielens,
A. G. G. M., Hollenbach, D. J., \& McKee, C. F. 1994, \apj, 433, 797
\bibitem[{Khachikian \& Weedman}(1974)]{KW74} Khachikian, E. Y. \&
Weedman, D. W. 1974, \apj, 192, 581
\bibitem[{Kishimoto}(1999)]{Kishimoto99} Kishimoto, M. 1999, \apj,
518, 676
\bibitem[{K\"onigl \& Kartje}(1994)]{KK94} K\"onigl, A. \& Kartje,
J. F. 1994, \apj, 434, 446
\bibitem[{Krolik \& Lepp}(1989)]{KL89} Krolik, J. H. \& Lepp,
S. 1989, \apj, 347, 179
\bibitem[{Lawrence}(1991)]{Lawrence91} Lawrence, A. 1991, \mnras,
252, 586
\bibitem[{Maloney, Hollenbach, \& Tielens}(1996)]{MHT96} Maloney,
P. R., Hollenbach, D. J., \& Tielens, A. G. G. M. 1996, \apj, 466, 561
\bibitem[{Maoz \& McKee}(1998)]{MM98} Maoz, E. \& McKee, C. F. 1998,
\apj, 494, 218
\bibitem[{Mathis, Rumpl, \& Nordsieck}(1977)]{MRN77} Mathis, J. S., Rumpl, W., \& Nordsieck, K. H. 1977, \apj, 217, 425
\apj, 494, 218
\bibitem[{Matt et al.}(1997)]{Metal97} Matt, G. et al. 1997,
\aap, 325, L13 
\bibitem[{Miller, Goodrich, \& Mathews}(1991)]{MGM91} Miller, J. S.,
Goodrich, R. W., \& Mathews, W. G. 1991, \apj, 378, 47
\bibitem[{Miyoshi et al.}(1995)]{Miyoshi95} Miyoshi, M., Moran, J.,
Herrnstein, J., Greenhill, L., Nakai, N., Diamond, P., \& Inoue,
M. 1995, \nat, 373, 127
\bibitem[{Murayama \& Taniguchi}(1997)]{MT97} Murayama, T. \&
Taniguchi, Y. 1997, \pasj, 49, L13
\bibitem[{Muxlow et al.}(1996)]{MPHGA96} Muxlow, T. W. B., Pedlar,
A., Holloway, A., Gallimore, J. F., \& Antonucci, R. R. J. 1996,
\mnras, 278, 854
\bibitem[{Nakai et al.}(1995)]{NIMMH95} Nakai, N., Inoue, M.,
Miyazawa, K., Miyoshi, M., \& Hall, P. 1995, \pasj, 47, 771
\bibitem[{Nakai, Inoue, \& Miyoshi}(1993)]{NIM93} Nakai, N., Inoue,
M., \& Miyoshi, M. 1993, \nat, 361, 45
\bibitem[{Nelson}(1996)]{Nelson96} Nelson, B. O. 1996, \apj, 465, L87
\bibitem[{Nenkova, Ivezic, \& Elitzur}(1999)]{NIE99} Nenkova, M.,
Ivezic, Z., \& Elitzur, M. 1999, Workshop on Thermal Emission
Spectroscopy and Analysis of Dust, Disks, and Regoliths, 
Houston, Texas, abstract no. 3020
\bibitem[{Netzer \& Peterson}(1997)]{NP97p85} Netzer, H., \& Peterson,
B. M. 1997, Astronomical Time Series, ed. D. Maoz, A. Sternberg, \& E. M. Leibowitz (Cambridge: Cambridge University
Press), 85
\bibitem[{Neufeld}(2000)]{Neufeld00} Neufeld, D. A. 2000, \apjl, 542,
L99
\bibitem[{Neufeld \& Kaufman}(1993)]{NK93} Neufeld, D. A., \& Kaufman, M. J. 1993, \apj, 263, 272
\bibitem[{Neufeld \& Maloney}(1995)]{NM95} Neufeld, D. A. \&
Maloney, P. R. 1995, \apj, 447, L17
\bibitem[{Neufeld, Maloney, \& Conger}(1994)]{NMC94} Neufeld, D. A.,
Maloney, P. R., \& Conger, S. 1994, \apj, 436, L127
\bibitem[{Ney \& Hatfield}(1978)]{NH78} Ney, E. P. \& Hatfield,
B. F. 1978, \apjl, 219, l111  
\bibitem[{Ott et al.}(1994)]{OWQKSSH94} Ott, M., Witzel, A.,
Quirrenbach, A., Krichbaum, T. P., Standke, K. J., Schalinski, C. J.,
\& Hummel, C. A. 1994, \aap, 284, 331
\bibitem[{Pier et al.}(1994)]{spectrum} Pier, E. A., Antonucci, R.,
Hurt, T., Kriss, G., \& Krolik, J. H. 1994, \apj, 428, 124
\bibitem[{Pier \& Krolik}(1993)]{PK93} Pier, E. A. \& Krolik,
J. 1993, \apj, 418, 99
\bibitem[{Press et al.}(1992)]
{numericalrecipes} Press, W. H., Teukolsky, S. A., Vetterling, W. T.,
\& Flannery, B. P. 1992, Numerical Recipes in C, Second Edition
(Cambridge: Cambridge University Press)
Catalog (Cambridge: Cambridge University Press)
\bibitem[{Reid \& Moran}(1988)]{RM88} Reid, M. J., \& Moran,
J. M. 1988, in Galactic and Extragalactic Radio Astronomy,
eds. G. L. Verschuur \& K. I. Kellermann (Berlin: Springer-Verlag),
255 
\bibitem[{Richer et al.}(1999)]{RSBC99} Richer, J., Shepherd, D.,
Bachiller, R., \& Churchwell, E. 1999, Protostars \& Planets IV, ed.
V. Mannings, A. Boss, \& S. Russell (Tucson: The University of Arizona
Press)
\bibitem[{Roy et al.}(1998)]{RCWU98} Roy, A. L., Colbert, E. J. M.,
Wilson, A. S., \& Ulvestad, J. S. 1998, \apj, 504, 147
\bibitem[{Salpeter}(1977)]{Salpeter77} Salpeter, E. E. 1977, \araa, 15,
267
\bibitem[{Seyfert}(1943)]{Seyfert43} Seyfert, C. K. 1943, \apj, 97,
28
\bibitem[{Tarter \& Welch}(1986)]{TW86} Tarter, J. C. \& Welch,
W. J. 1986, \apj, 305, 467
\bibitem[{Tully}(1988)]{Tully88} Tully, R.B. 1988, Nearby Galaxies
Catalog (Cambridge: Cambridge University Press)
\bibitem[{Voit}(1991)]{Voit91} Voit, G. M. 1991, \apj, 379, 122
\bibitem[{Watson \& Wallin}(1994)]{WW94} Watson, W. D. \& Wallin,
B. K. 1994, \apjl, 432, L35
\bibitem[{Watson \& Wallin}(1997)]{WW97} Watson, W. D. \& Wallin,
B. K. 1997, \apjl, 476, 685
\bibitem[{Weinberger, Neugebauer, \& Matthews}(1999)]{WNM99}
Weinberger, A. J., Neugebauer, G., \& Matthews, K. 1999, \aj, 117,
2748
\end{thebibliography}
